\documentstyle[prb,aps,eqsecnum,epsfig,floats]{revtex}

\newcommand{\bfsigma}{\sigma\hspace{-0.58em}\sigma\hspace{-0.58em}\sigma\hspace{-0.58em}\sigma}

\begin{document}
\draft
\twocolumn[{\hsize\textwidth\columnwidth\hsize\csname@twocolumnfalse%
\endcsname

\title{NMR properties of a one-dimensional Cu-O model}
\author{T. Becker, M. Gabay, and T. Giamarchi}
\address{Laboratoire de Physique des Solides, CNRS UMR 85002, Universit\'e
Paris-Sud, 91405 Orsay, France}
\date{July 17, 2000}
\maketitle

\begin{abstract}
We obtain the Knight shifts and the relaxation rates related to the Fermi
 contact interaction term for a one-dimensional Cu-O model using
 bosonization technique. We consider the small interaction limit
 at half-filling and away from half-filling.  In this framework we predict
that the antiferromagnetic contribution to the relaxation rate of the
nuclear oxygen spin is completely suppressed even away from half-filling,
when the temperature is low enough. In the 
strong interaction limit at half-filling we compute the
effective Fermi contact interaction performing a Gutzwiller
 projection. Both limits suggest that the one-dimensional versions of the 
Mila-Rice and of the Shastry scenarios of transferred hyperfine couplings 
which were proposed to explain the NMR measurements for High-$\rm T_c$ 
cuprates fail in a one-dimensional situation.
\end{abstract}

\pacs{74.72-h, 76.20.+q, 76.60-k}}]
\narrowtext

\section{Introduction}

One of the challenging characteristics of the cuprate materials is the
importance of magnetic fluctuations and their effect on normal state
transport and on superconductivity. This feature is naturally present 
in theoretical approaches emphasizing strong interactions in 2D. The 
role of magnetism can also be assessed in other scenarios promoting marginal
or nearly antiferromagnetic Fermi liquid (NAFL) behavior. The NAFL
framework has been used by Millis, Monien, and Pines \cite{millis_monien_pines_NAFL}
to discuss NMR experiments: on the basis of the Mila-Rice
\cite{mila_rice_trans_hyp_coupling} and Shastry \cite{shastry_trans_hyp_coupling} 
local terms describing spin fluctuations induced by the hyperfine
interactions, these authors were able to compute various Knight shifts and
nuclear relaxation times. However, there are critical remarks and further
details to this theory,\cite{anderson_against_NAFL,zha_nmr_critical_reexamination,millis_nmr_critical_reexamination,bulut_trans_hyp_coupling_rpa}
and also other theories introduced in order to interpret the behavior of the
various Knight shifts and nuclear relaxation times in high-$\rm T_c$
superconductors, which we do not discuss here. \cite{varma_nmr_in_CuO}

In this paper, we will follow the basic assumption of Mila and Rice that it
is necessary to include a sizeable isotropic hyperfine interaction term 
to fit the data of NMR experiments in High-$\rm T_c$ cuprates. Thus in the 
following we will focus our interest on this contribution which 
comes mainly from the Fermi contact interaction between a nucleus and 
its surrounding partially filled s orbitals,
namely the 4s orbital for copper and the 3s orbital for oxygen.
 Experiments have shown that the magnetic properties are described by a
single-spin component model. In the weak interaction limit,
this single-spin degree of freedom could be associated with the strongly 
hybridized Cu--3d---O--2p anti-bonding band,
\cite{millis_nmr_critical_reexamination,bulut_trans_hyp_coupling_rpa} 
whereas in the strong interaction limit it is associated with the
 nearly localized Cu-3d spin. \cite{mila_rice_trans_hyp_coupling,shastry_trans_hyp_coupling,bulut_trans_hyp_coupling_rpa}
Within the local picture, the contrasting NMR behavior
seen on the Cu and O sites arises from their different hyperfine 
form factors. These are nothing but the Fourier transforms of the Fermi
contact interaction terms approximated by a sum over surrounding
localized next-neighbor Cu--3d spins.

One of the (many) complications concerning the physics of cuprate materials
is that there is still no consensus about what should be the correct 
theoretical approach to treat correlations in 2D: Is the ground state 
Fermi liquid or non-Fermi liquid like?
Can one treat interactions perturbatively, or is it more appropriate
to treat kinetic terms as corrections in the strongly interacting limit?
In the latter category, working out a consistent treatment of the non-double occupancy
constraint is still an open issue.

By contrast, one-dimensional systems offer a perfect testing ground for the
study of magnetic fluctuations. Since in one dimension it is
possible to treat correlation effects properly both in the limit of
weak and strong interactions, such models allow to
compute explicitly the dependence of the relaxation. This allows
to get some feeling for the effects of doping. In addition to the
insight that such study allows to gain for higher dimensional
models, there are explicit realizations of one-dimensional systems,
such as the Bechgaard salts or copper germanate compounds.\cite{jerome_revue_1d,bourbonnais_normal_phase_of_1D_org_sc}

For these reasons, we choose to investigate hyperfine interactions 
in the one-dimensional version of the Cu-O model. This allows us to extract
form factors both in the insulating and in the doped regime, without assuming a
specific form of the Fermi contact interaction term.
We can then
compare the exact results with the predictions that the standard
approximation schemes used in two dimensions would give in the one-dimensional situation.

The paper is organized as follows. In section~\ref{definition_models}, we
introduce the model in one dimension, as well as the three different
approximations used to describe the magnetic relaxation processes
related to the Fermi contact interaction term
, namely that due to Mila-Rice, \cite{mila_rice_trans_hyp_coupling} to Shastry,
\cite{shastry_trans_hyp_coupling} and to Bulut. \cite{bulut_trans_hyp_coupling_rpa} In
section~\ref{weak_interaction_limit}, we solve the full model for weak
interactions as compared to the bandwidth. We obtain spin-spin correlation functions and
discuss the Knight shift $K$ and the relaxation rate
$1/T_1$ in detail. We compare our results with the prediction of the Bulut
model, which is applicable for weak interactions.
Section~\ref{strong_interaction_limit}
solves the problem in the opposite limit of very strong
interactions using a Gutzwiller projection eliminating double
occupation on the copper sites. We again compare our results with
the one-dimensional extrapolation of the Mila-Rice and of the Shastry
approximations. A general discussion of our results is presented
in section~\ref{discussion and
perspectives}. Since NMR data on organic\cite{thank_jerome_wzietek} and
inorganic quasi-1D compounds seem to give an essentially isotropic relaxation
rate, the body of the paper mostly focuses on the isotropic contribution to
the hyperfine interaction. Yet, for the sake of completeness and in
view of the fact that $K$ can be anisotropic (see below in section V) we
discuss the effect of anisotropic hyperfine terms in Appendix~\ref{anistropic_hyperfin}: these terms only modify prefactors in the 
expressions of $1/T_1$ and of $K$. Appendix~\ref{sine-gordon-model} and \ref{local_states_for_u_a} offer details of our calculations.

\section{Definition of the models}\label{definition_models}

\subsection{The four-band model}

\begin{figure}
\centerline{\epsfig{file=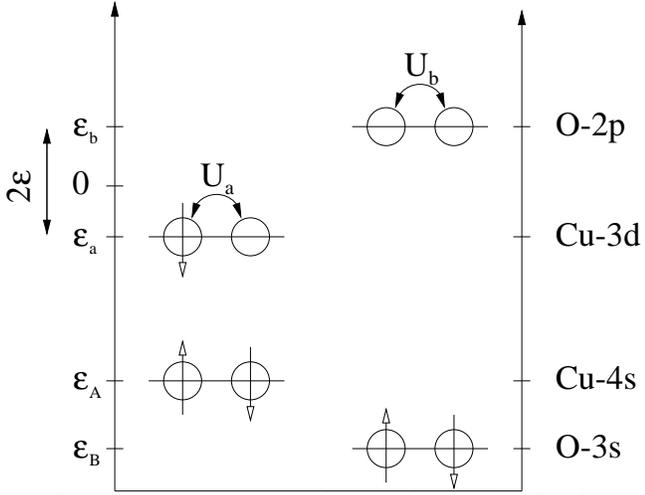,angle=0,width=8.6cm}}
\caption{Energy of the various orbitals. The Cu--3d and O--2p  
(resp. $a$ and $b$) orbitals are the relevant ones to describe 
the electronic degrees of freedom.
The Cu--4s and O--3s  (resp. $A$ and $B$) should be kept to
describe the coupling to the nuclear spin via a Fermi contact interaction.}
\label{fig:orbitdef}
\end{figure}

We consider a system with two different atoms per unit cell (Cu and O). In
order to describe the ground state properties, we take into
account the 3d and 2p orbitals on Cu and O respectively. The related hole states are
represented in Fig.~\ref{fig:orbitdef} and denoted by $a$ and $b$
in the following.
Since the coupling to the nuclear spin via the Fermi contact interaction
occurs only for partially filled s orbitals, we also have to retain the
Cu--4s and O--3s shells
(denoted by $A$ and $B$) to correctly obtain the desired NMR properties. The Hamiltonian describing the system can thus be
written as
\begin{equation} \label{eq:hamcomp}
H = H_0 + H_S + H_N\;,
\end{equation}
where $H_0$ contains the electronically relevant orbitals $a$ and
$b$. $H_S$ describes the coupling of orbitals $a$ and $b$ to orbitals $A$
and $B$.  As indicated in Fig.~\ref{fig:orbitdef}, the orbitals $A$ and $B$ are basically filled  and produce only small
corrections to the electronic term represented by $H_0$, so that we will treat
$H_S$ as a perturbation.
Finally, $H_N$ describes the coupling of the orbitals $A$ and $B$ 
to the nuclear spins and will be treated as a small -- time dependent -- perturbation in linear response.

\begin{figure}
\centerline{\epsfig{file=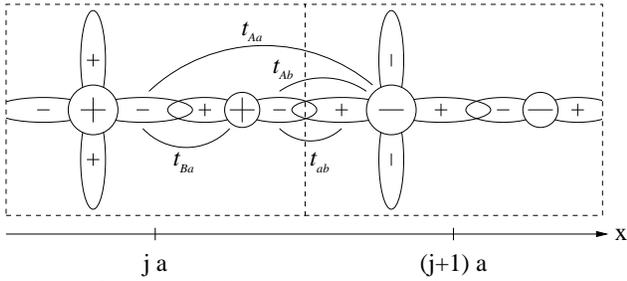,angle=0,width=8.6cm}}
\caption{Cu--3d and Cu--4s orbitals as well as the O--2p and O--3s orbitals in the unit cells. The chosen signs
of the wave functions determine the phase convention for the Hamiltonian and the
signs of the various tight-binding parameters $t_{\eta\eta^\prime}$.}
\label{fig:phaseconv}
\end{figure}

The main contribution, $H_0$, is given by
\begin{equation}\label{eq:ham0}
H_0= H_T + H_U\;,
\end{equation}
where
\begin{eqnarray}\label{eq:hamT}
H_T &=& \sum_j\epsilon_a n_{aj} + \epsilon_b n_{bj}\nonumber\\
    &-& \sum_{j\sigma} t_{ab} \left[a^\dagger_{j\sigma} (b_{j\sigma} +
  b_{j-1,\sigma}) + \text{h.c.} \right]
\\      \label{eq:hamU}
H_U &=& \sum_j U_a n_{aj\uparrow } n_{aj\downarrow}
      + \sum_j U_b n_{bj\uparrow} n_{bj\downarrow}\;.
\end{eqnarray}
Here $t_{ab}$ describes the hopping between the Cu-3d and the O-2p
orbitals with the phase conventions shown in Fig.~\ref{fig:phaseconv}.
 $U_a$ and $U_b$ are the local repulsions on the copper and oxygen sites as
shown in Fig.~\ref{fig:orbitdef}. The Coulomb repulsions $U_A$ and $U_B$ can
be
ignored assuming that energy cost considerations discourage processes in which two holes are
excited in one s orbital. A nearest neighbor interaction $U_{ab}$ could
also be added to the model to generate a phase transition to a
superconducting phase, \cite{stechel_CuO_model} but here our study deals with the vicinity of
the half-filled case, i.e. near the antiferromagnetic phase, and we will
ignore $U_{ab}$.

The coupling between the orbitals $a,b$ and $A,B$ reads
\begin{eqnarray} \label{eq:hamS}
H_S & = & \sum_j \epsilon_A n_{Aj} + \epsilon_B n_{Bj}\nonumber \\
    & + &  \sum_{j\sigma} t_{Ba}
                \left[B^\dagger_{j\sigma}(a_{j+1,\sigma} - a_{j\sigma})
                      + \text{h.c.} \right] \nonumber \\
    & + &   \sum_{j\sigma} t_{Ab}
                \left[A^\dagger_{j\sigma}(b_{j\sigma} + b_{j-1,\sigma})
                      + \text{h.c.} \right] \nonumber \\
    & + &   \sum_{j\sigma} t_{Aa}
                \left[A^\dagger_{j\sigma}(a_{j+1,\sigma} + a_{j-1,\sigma})
                      + \text{h.c.} \right]
\end{eqnarray}
with the phase conventions of Fig.~\ref{fig:phaseconv}. The density
operators $n_{\eta j}$ in (\ref{eq:hamT}), (\ref{eq:hamU}), and (\ref{eq:hamS})
are the standard ones
\begin{equation} \label{eq:density}
n_{\eta j} = \sum_\sigma n_{\eta j\sigma}
            =\sum_\sigma \eta^\dagger_{j\sigma}\eta_{j\sigma}\;,
\end{equation}
where $\eta = a,b,A,B$.

Finally, the isotropic coupling to the nuclear spins (${\bf I}$ for the
copper atom and ${\bf J}$ for the oxygen atom) is given by
\begin{equation} \label{eq:hamN}
H_N  =  \sum_j C_A {\bf I}_{j} {\bf S}_{Aj}
            +  C_B {\bf J}_{j} {\bf S}_{Bj}\;,
\end{equation}
where similarly to (\ref{eq:density}) the spin operators are given by
\begin{equation} \label{eq:spindensity}
{\bf S}_{\eta j} = \frac{1}{2}\sum_{\sigma_1\sigma_2}
\eta^\dagger_{j\sigma_1} {\bfsigma}_{\sigma_1\sigma_2} \eta_{j\sigma_2}\;.
\end{equation}
The coupling constant for the electron-nuclear interaction is given by $C_\eta
=(8 \pi / 3) |\psi_\eta (0) |^2 \gamma_n g \mu_B \hbar$, which is
proportional to the local hole density for the respective s orbitals at the
origin. $\sigma_{\sigma_1\sigma_2}$ are the Pauli matrices.

\subsection{Reduced models}

In 1D, the Hamiltonian (\ref{eq:hamcomp}) can explicitly be expressed
in terms of bose fields. This allows for a full treatment, in which
$H_0$, $H_S$, and $H_N$ are all treated on an equal footing. This is
the route we follow in section~\ref{weak_interaction_limit}. In higher
 dimension there is
still no solution of the fully interacting problem. So various
approximation schemes have been devised and applied to each of the pieces
of $H$ separately. These lead to effective "nuclear" Hamiltonians : the
Mila-Rice and the Shastry model for strong interactions and the Bulut model for weak interactions. Typically, the
analysis by Mila-Rice and Shastry starts from a partially projected Hamiltonian
\begin{equation}
H=\hat{P} H_0 \hat{P}  + H_S + H_N\;.
\end{equation}
The first part gives the t-J model or at half-filling the Heisenberg model, which contains the dynamics related to the Cu-3d and O-2p
orbitals. The second part contains the unprojected degrees of
freedom related to the Cu-4s and O-3s orbitals, as well as the electron-nuclear
interaction part. Further approximations for $H_S$ and $H_N$ lead to the Mila-Rice model
or to the Shastry model (see below).
In the weak interaction limit, Bulut et al. have proposed an RPA treatment of
$H_0$ in combination with the effective electron-nuclear interaction term of
Mila-Rice and Shastry.

Before we turn to the full solution of the
model, let us review the main features of such approximations when
applied to our one-dimensional system. This will allow us to contrast the
predictions of the 1D version of these three models and the results obtained
for the four-band model, which may provide some clue to the validity
of these approaches for strongly correlated systems.

\subsubsection{The Mila-Rice model for strong interaction}

The model defined in (\ref{eq:hamcomp}) is approximated by
\begin{equation}
H \simeq H_0^{Mi} + H_N^{Mi}\;. \label{eq:hamcompMR}
\end{equation}
$H_0^{Mi}$ is the approximation for (\ref{eq:ham0}) and
denotes, at half-filling, a Heisenberg model for local Cu-3d spins generated by
\begin{equation}
 H_0^{Mi}=\hat{P}\big[ H_0(U_b=0)\big] \hat{P}\;,
\end{equation}
where  $\hat{P}$ is the Gutzwiller projection operator which
prohibits doubly occupied Cu-3d states. The
additional unprojected part (\ref{eq:hamS}) with $t_{Aa}=t_{Ba}=0$ and the electron-nuclear interaction part
(\ref{eq:hamN}) are approximated by
\begin{equation}\label{eq:hamNMilaRice}
H_N^{Mi}=\sum_j C_A{\bf I}_{j}{\bf S}_{Aj}^{Mi}\;,
\end{equation}
where
\begin{equation}\label{eq:copper_spin_approx_mila_rice}
{\bf S}_{Aj}^{Mi} = F_{Aa}^{Mi} \left( {\bf S}_{a,j-1}+ {\bf S}_{a,j+1}\right)\;.
\end{equation}
Thus ${\bf S}_{Aj}^{Mi}$ is the Mila-Rice approximation for the original
spin ${\bf S}_{Aj}$ used to explain the NMR experiments measured on the copper sites. $F_{Aa}^{Mi}=|\lambda_{Aa}^{Mi}|^2$ denotes the
effective overlap between one Cu-4s spin with a neighboring Cu-3d
spin. In the Mila-Rice model only hopping processes via the O-2p orbitals
are included, whereas the direct hopping between Cu-3d and Cu-4s
orbitals is ignored. In Ref.~\onlinecite{mila_rice_trans_hyp_coupling} Mila and Rice perform a quantum chemical
analysis without including interaction effects, so we will do the same and
propose for the amplitudes
\begin{equation}
\lambda_{Aa}^{Mi}=-\frac{t_{Aa}t_{ab}}{(\epsilon_a-\epsilon_A)(\epsilon_a-\epsilon_b)}\;.
\end{equation}
This result is obtained by a projection in real space of a Cu-4s orbital
onto a neighboring Cu-3d orbital for $U_a=0$.
The Fourier transform of
(\ref{eq:copper_spin_approx_mila_rice}) is given by
\begin{equation}
{\bf S}_{Ap}^{Mi} = F_{Aa}^{Mi}(p)  {\bf S}_{ap}
\end{equation}
with the Mila-Rice form factor
\begin{equation} \label{eq:formfactor_copper_mila_rice}
F_{Aa}^{Mi}(p)=2 F_{Aa}^{Mi} \cos(pa)\;.
\end{equation}

\subsubsection{The Shastry model for strong interaction}

The approximation proposed by Shastry in
Ref.~\onlinecite{shastry_trans_hyp_coupling} is given by
\begin{equation}
H \simeq H_0^{Sh}+ H_N^{Sh} \label{eq:hamcompSh}\;,
\end{equation}
where
\begin{equation}
 H_0^{Sh}=\hat{P}\big[ H_0(U_b=0,U_a=\infty)\big]\hat{P}
\end{equation}
leads to the Heisenberg model at half-filling and to the t-J
model for a doped system with strong repulsion on the copper sites. The electron-nuclear interaction part reads
\begin{equation}\label{eq:hamNShastry}
H_N^{Sh}=\sum_j C_A{\bf I}_{j}{\bf S}_{Aj}^{Sh}
        +  C_B{\bf J}_{j}{\bf S}_{Bj}^{Sh}\;,
\end{equation}
where the spins ${\bf S}_{\eta j}^{Sh}$ are approximated by a linear
combination of unprojected Cu-3d orbitals
\begin{eqnarray}
{\bf S}_{Aj}^{Sh} & = & F_{Aa}^{Sh} \left(
{\bf S}_{a,j-1}+ {\bf S}_{a,j+1}\right)
\nonumber\\
 &+&F_{Ab}^{Sh} \left({\bf S}_{bj}+ {\bf S}_{b,j-1}\right)
\nonumber\\
&\cong &F_{Aa}^{Sh} \left(
{\bf S}_{a,j-1}+ {\bf S}_{a,j+1}\right)
\label{eq:copper_spin_approx_shastry}
\end{eqnarray}
and
\begin{equation}
{\bf S}_{Bj}^{Sh} = F_{Ba}^{Sh} \left( {\bf S}_{aj}+ {\bf S}_{a,j+1}\right)\label{eq:oxygen_spin_approx_shastry}
\end{equation}
with the coefficients
\begin{eqnarray}
 F_{Aa}^{Sh}=|\lambda_{Aa}^{Sh}|^2 &=& \left(
\frac{t_{Aa}}{\epsilon_A-\epsilon_a} \right)^2\label{eq:coef_shastry_Aa}
\\
 F_{Ab}^{Sh}=|\lambda_{Ab}^{Sh}|^2 &=&\left(\frac{t_{Ab}}{\epsilon_A-\epsilon_b} \right)^2\label{eq:coef_shastry_Ab}
\\
 F_{Ba}^{Sh}=|\lambda_{Ba}^{Sh}|^2 &=&\left(\frac{t_{Ba}}{\epsilon_B-\epsilon_a} \right)^2\;.\label{eq:coef_shastry_Ba}
\end{eqnarray}
For finite doping, it is assumed in
Ref.~\onlinecite{shastry_trans_hyp_coupling} that the spin degrees of freedom related to ${\bf S}_{bj}$ are quenched in a Zhang-Rice singlet.
\cite{zhang_rice_singlet} This assumption justifies the second
approximation done in (\ref{eq:copper_spin_approx_shastry}). Further, Shastry includes
only the direct couplings (\ref{eq:coef_shastry_Aa}) up to second order
proportional to
$t_{Aa}^2$  for the relaxation of the nuclear copper spin and ignores the
fourth order contributions proportional to $t_{ab}^2t_{Ab}^2$ via the O-2p
orbital as proposed by Mila and Rice. The Fourier transform of the
approximated spins ${\bf S}_{\eta j}^{Sh}$ reads
\begin{equation}
{\bf S}_{\eta p}^{Sh} = 2 F_{\eta a}^{Sh}(p) {\bf S}_{ap}
\end{equation}
with
\begin{eqnarray}
F_{Aa}^{Sh}(p)& = & 2 F_{Aa}^{Sh}\cos(pa)
\label{eq:formfactor_copper_shastry}
\\
F_{Ba}^{Sh}(p)& = & 2 F_{Ba}^{Sh}\cos(pa/2)\;.
\label{eq:formfactor_oxygen_shastry}
\end{eqnarray}
For the uniform contribution ($p \sim 0$) all form factors are finite,
but for the antiferromagnetic wave vector ($p \sim \pi/a$) the form factor
vanishes for the oxygen sites, whereas it stays finite for the copper sites.
Thus (\ref{eq:formfactor_copper_mila_rice}),
(\ref{eq:formfactor_copper_shastry}), and (\ref{eq:formfactor_oxygen_shastry}) are the one-dimensional analogs of
the NMR Mila-Rice and Shastry form factors for High-${\rm T_c}$ cuprates.

\subsubsection{The Bulut model for weak interaction}

In Ref.~\onlinecite{bulut_trans_hyp_coupling_rpa}, Bulut et al. used a weak
 interaction RPA calculation combined with an electron-nuclear interaction
 as proposed by Mila-Rice and Shastry to compute NMR related quantities for the
 High-$\rm T_c$ cuprates. The one-dimensional analog reads
\begin{equation} \label{eq:hamcompBu}
H \simeq H_0^{Bu} + H_N^{Bu}\;.
\end{equation}
In two dimensions, $H_0^{Bu}$ is obtained by applying the RPA method to the 
2D version of (\ref{eq:ham0}) with $U_b=0$. In one
dimension, we can treat all the interaction terms ($U_a,U_b$) of the original
Hamiltonian $H_0$ by means of bosonization and of
renormalization-group theory. Finally, the electron-nuclear interaction term  $H_N^{Bu}$ is given in analogy to
(\ref{eq:hamNShastry}) replacing the approximated spins by 
\begin{eqnarray}
{\bf S}_{Aj}^{Bu} & = & F_{Aa}^{Bu} \left(
{\bf S}_{a,j-1}+ {\bf S}_{a,j+1}\right)
\label{copper_spin_approx_bulut}
\\
{\bf S}_{Bj}^{Bu} & = & F_{Ba}^{Bu} \left( {\bf S}_{aj}+ {\bf
S}_{a,j+1}\right)\;.
\label{eq:oxygen_spin_approx_bulut}
\end{eqnarray}
In Ref.~\onlinecite{bulut_trans_hyp_coupling_rpa}, the parameters $F_{\eta a}^{Bu}$ are
undefined and could in general include all possible overlaps of the Cu-4s
and O-3s orbitals with the Cu-3d orbitals in the sense of Mila-Rice and
Shastry. The coefficient for the oxygen will be
\begin{equation}
 F_{Ba}^{Bu}=|\lambda_{Ba}^{Bu}|^2 =|\lambda_{Ba}^{Sh}|^2\;,
\end{equation}
and is thus the same as that proposed by
Shastry, whereas the coefficient for copper
\begin{equation}
 F_{Aa}^{Bu}=|\lambda_{Aa}^{Bu}|^2=|\lambda_{Aa}^{Sh}+\lambda_{Aa}^{Mi}|^2
\end{equation}
includes additional combined terms of third order proportional to $t_{Aa}t_{Ab}t_{ab}$, which are absent in the
Mila-Rice and the Shastry model.
The form factors correspond to
\begin{eqnarray}
F_{Aa}^{Bu}(p)&=&2 F_{Aa}^{Bu}\cos(pa)
\label{eq:formfactor_copper_bulut}
\\
F_{Ba}^{Bu}(p)&=&2 F_{Ba}^{Bu}\cos(pa/2)\;.
\label{eq:formfactor_oxygen_bulut}
\end{eqnarray}

\section{The weak interaction limit}\label{weak_interaction_limit}

Let us now solve the full model (\ref{eq:hamcomp}) when interactions are
weak compared to the bandwidth. This allows us to use the
bosonization technique for treating interactions in the undoped as well as in
the doped case.

\subsection{NMR properties of the four-band model}
\label{NMR_properties_general}

\subsubsection{Reduction to an effective single-band Hamiltonian}
\label{reduction_eff_single_band}

Instead of working with the basis $a,b$ it is more convenient to
diagonalize (\ref{eq:hamT}) within a unit cell, and to introduce the
bonding and anti-bonding bands. Using the transformation
\begin{eqnarray} \label{eq:bonding}
a_{k\sigma} & = & \left[\cos(\gamma_k) \alpha_{k\sigma}
                - \sin(\gamma_k) \beta_{k\sigma}\right] e^{-i \frac{k a}{2}}
\nonumber \\
b_{k\sigma} & = & \sin(\gamma_k) \alpha_{k\sigma}
                + \cos(\gamma_k) \beta_{k\sigma}
\end{eqnarray}
with
\begin{equation}
\tan(2\gamma_k) = \frac{2 t_{ab}}{\epsilon} \cos( k a/2 ) \quad
,\quad \gamma_{k} \in \left[ 0, \frac{\pi}{4} \right[ \\
\end{equation}
the kinetic energy (\ref{eq:hamT}) becomes
\begin{equation}
H_T =  \sum_{k\sigma} \left[
       \epsilon_\alpha(k) \alpha^\dagger_{k\sigma}\alpha_{k\sigma}
     + \epsilon_\beta(k) \beta^\dagger_{k\sigma}\beta_{k\sigma} \right]\;,
\end{equation}
where the state $| \alpha_{k\sigma} \rangle$ refers to the lower Hubbard band
with energy $\epsilon_\alpha (k)= -\epsilon /\cos(2\gamma_k)$, and the state
$| \beta_{k\sigma} \rangle$ to the upper one with energy $\epsilon_\beta
(k)=\epsilon /\cos(2\gamma_k)$. In the absence of interactions the chemical potential $\mu$
lies in the $\alpha$-band both for the undoped and for the doped system,
and one can ignore the $\beta$-band, which is at least
$\epsilon_b-\epsilon_a = 2\epsilon$ higher in
energy.
The same property holds when the interaction terms (\ref{eq:hamU}) are added
to (\ref{eq:hamT}), given that in the weak coupling limit $U_b,U_a \ll
2t_{ab}^2 / \epsilon$. Correlation effects in 1D will strongly affect the $\alpha$-band
states, thus in the following we
ignore the terms containing $\beta$-operators when substituting
(\ref{eq:bonding}) into (\ref{eq:hamU}).

Substituting (\ref{eq:bonding}) into (\ref{eq:hamS}) and performing a
first order perturbation theory with respect to $H_S$, all operators in
(\ref{eq:hamcomp}) can be written as
\begin{equation}\label{eq:proj}
\eta_{k\sigma} \cong \lambda_{\eta\alpha}(k)\alpha_{k\sigma}
\end{equation}
with
\begin{eqnarray} \label{eq:coefproj}
\lambda_{a\alpha}(k) & = & \cos(\gamma_k) e^{-i\frac{k a}{2}} \nonumber\\
\lambda_{b\alpha}(k) & = & \sin(\gamma_k) \nonumber \\
\lambda_{A\alpha}(k) & = & \frac{2 t_{Aa} \cos(k a)}
    {\epsilon_\alpha(k)-\epsilon_A}\cos(\gamma_k)e^{-i\frac{k a}{2}} \nonumber \\
    & + & \frac{2 t_{Ab} \cos(k a/2)}
            {\epsilon_\alpha(k)-\epsilon_A}\sin(\gamma_k)e^{-i\frac{k a}{2}} \nonumber \\
\lambda_{B\alpha}(k) & = & \frac{2 i t_{Ba} \sin(k a/2)}
            {\epsilon_\alpha(k)- \epsilon_B} \cos(\gamma_k)\;.
\end{eqnarray}
Here we have assumed an unperturbed ground state $|\alpha^{(1)}\rangle\cong
|\alpha\rangle$. Thus (\ref{eq:proj}) implies that (\ref{eq:hamcomp}) reduces to an effective single-band Hamiltonian.
\subsubsection{The Continuum limit}

\begin{figure}
\centerline{\epsfig{file=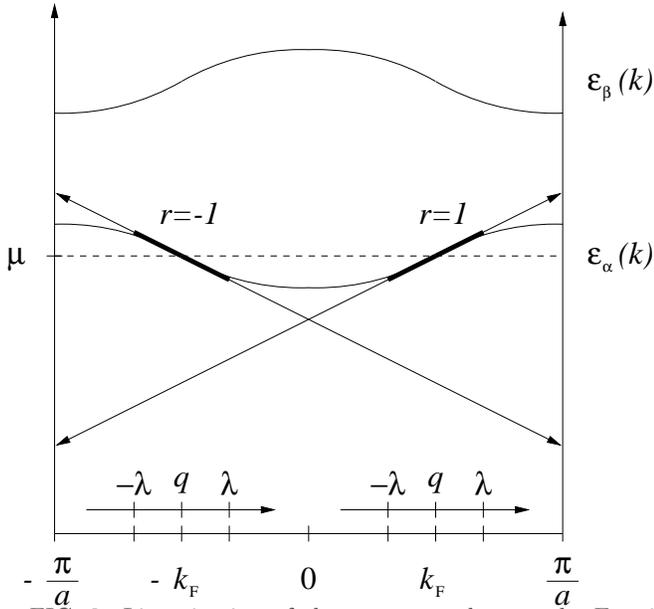,angle=0,width=8.6cm}}
\caption{Linearization of the spectrum close to the Fermi points. The
momentum $k$ is replaced by $r k_F+q$, where $r=\pm$ denotes the two possible
directions. $\lambda >|q|$ is an ultraviolet cutoff of the order of the bandwidth
and the Fermi velocity is $v_F = \partial_k \epsilon_\alpha(k) |_{k_F}$.}
\label{fig:linearize}
\end{figure}

We can now use the standard techniques in order to treat interacting
one-dimensional systems. Restricting ourselves to the low energy
physics regime we make the usual approximation valid for 1D systems, i.e we
linearize the spectrum close to the
Fermi points, as shown in Fig.~\ref{fig:linearize}.
Then the Hamiltonian (\ref{eq:ham0}) is reduced to
\begin{eqnarray}
H_T & = & \sum_{r=\pm ,q,\sigma} r v_F q \alpha^\dagger_{rq\sigma}
\alpha_{rq\sigma} \\ \label{eq:hamlowU}
H_U & = &\sum_{\bf r,q} \frac{U({\bf r})}{N}
            \alpha^\dagger_{r_1,q_1+q_3\uparrow}
            \alpha^\dagger_{r_2,q_2-q_3\downarrow}
            \alpha_{r_3,q_2\downarrow}
            \alpha_{r_4,q_1\uparrow}\;,
\end{eqnarray}
where ${\bf r}=(r_1,r_2,r_3,r_4)$ and  ${\bf q}=(q_1,q_2,q_3)$. $U({\bf r})$ parameterizes the repulsive interaction in the continuum
limit and is given in terms of
the standard notations as
\begin{eqnarray}
U(\pm,\pm,\pm,\pm) &\equiv& U_o\nonumber\\
U(\pm,\mp,\mp,\pm) &\equiv& U_o\nonumber\\
U(\pm,\mp,\pm,\mp) &\equiv& U_s\nonumber\\
U(\pm,\pm,\mp,\mp) &\equiv& U_c\;.
\end{eqnarray}
$U_o$ refers to the two forward scattering processes,  $U_s$ to the
backward scattering, and  $U_c$ to the umklapp scattering process that occurs
at half-filling.
The relation to the local repulsions defined in (\ref{eq:hamU}) is given by
\begin{eqnarray}
U_o & = & U_{b\alpha} + U_{a\alpha} > 0\nonumber \\
U_s & = & U_{b\alpha} + U_{a\alpha} > 0\nonumber \\
U_c & = & U_{b\alpha} - U_{a\alpha} < 0\;,
\end{eqnarray}
where $U_{\eta\alpha}$ is the
short notation for the projected Coulomb energies $U_\eta |\lambda_{\eta\alpha}(k_F)|^4$.

As usual for interacting one-dimensional systems, it is useful to
introduce a boson representation of the fermion operators, related to
the charge and spin density fluctuations.
Since the technique is standard, we only recall the main steps and refer the
reader to the literature. \cite{mattis_backscattering,luther_bosonisation,haldane_bosonisation,heidenreich_bosonisation} We rewrite the original
density operators in terms of a linear combination of charge ($\nu =c$)
and spin ($ \nu =s$) density operators for each branch
\begin{equation}
\rho_{r\sigma} = (\rho_{rc} + \sigma \rho_{rs})/\sqrt{2}\;.
\end{equation}
These density operators define the phase fields
\begin{eqnarray} \label{eq:phasefields}
\Phi_\nu (x) & = & - \frac{i\pi}{L}\sum_{r,q \neq 0}\frac{1}{q}
e^{-a |q|/2- i q x} \rho_{r\nu}\nonumber \\
\Theta_\nu (x) & = &  \frac{i\pi}{L}\sum_{r,q \neq 0}\frac{r}{q}
e^{-a |q|/2- i q x} \rho_{r\nu}\;.
\end{eqnarray}
All operators can be expressed in terms of the boson fields
(\ref{eq:phasefields}), and the fermion operator reads:
\begin{equation} \label{eq:fermion_as_boson}
\alpha_{r\sigma}(x) =  \frac{1}{\sqrt{2 \pi a}}
        e^{ir k_F x - \frac{i}{\sqrt{2}} \left[ r ( \Phi_c  + \sigma \Phi_s)
                        -(\Theta_c + \sigma \Theta_s )
                    \right]}\;.
\end{equation}
The complete Hamiltonian becomes
\begin{equation} \label{eq:boscomp}
H =(H_0^c + H_U^c) +(H_0^s + H_U^s)+ H_N\;,
\end{equation}
where 
\begin{equation} \label{eq:bosquad}
H_0^\nu  =  \int  \frac{dx}{2\pi}
            \left[(u_\nu K_\nu)(\pi \Pi_\nu)^2
            +   \left(\frac{u_\nu}{K_\nu}\right) (\partial_x \Phi_\nu)^2 \right]
\end{equation}
is a quadratic part containing
only charge or spin degrees of freedom (with $\nu=c,s$).  
In (\ref{eq:bosquad}), the variable $\Pi_\nu = \partial_x
\Theta_\nu$ is the momentum density conjugate to $\Phi_\nu$, and thus they respect
the commutation relation $[\Phi_\nu(x),\Pi_\nu(x^\prime)]=i\delta(x-x^\prime)$.
The interaction terms are given by
\begin{eqnarray} \label{eq:bosintc}
H_U^c & = &    \int dx \frac{2 a U_{b\alpha}}{(2\pi a)^2}
                \cos[\sqrt{8}\Phi_c - \delta x] \nonumber\\
      & - &    \int dx \frac{2 a U_{a\alpha}}{(2\pi a)^2}
                \cos[\sqrt{8}\Phi_c - \delta (x-a/2)] 
\\
H_U^s & = &     \int dx \frac{2 a U_s}{(2\pi a)^2}
                                \cos[\sqrt{8}\Phi_s]\;.
\end{eqnarray}
Here $\delta=4 k_F- 2 \pi /a$ is proportional
to the doping of the system with respect to the half-filled case shown in
Fig.~\ref{fig:orbitdef} (for which $k_F=\pi /2a$). Using this representation
we suppose to work with a fixed number of particles, since $k_F$ is directly
related to the filling.
Finally, the isotropic electron-nuclear interaction part could be written as
\begin{equation} \label{eq:bosintN}
H_N =      \int dx \left[
                C_{A} {\bf I}(x){\bf S}_{A}(x)
            +   C_{B} {\bf J}(x){\bf S}_{B}(x)
            \right]\;,
\end{equation}
where $H_N$ is the projection of (\ref{eq:hamN}) onto the $\alpha$-band
using  (\ref{eq:spindensity}) and (\ref{eq:proj}).
The projected spin operators ${\bf S}_\eta$ are expressed in terms of (\ref{eq:fermion_as_boson}). For example, the $z$-component of
the spin operator ${\bf S}_{\eta}$ can be
represented as a sum of $p \sim 0$ and $p \sim 2k_F$ components as
\begin{equation} \label{eq:spincomp}
S_{\eta}^z(x)=|\lambda_{\eta\alpha}(k_F)|^2[\bar{s}_{\alpha}(x) + \tilde{s}_{\eta\alpha}(x)]\;,
\end{equation}
where the non-oscillatory part is given by
\begin{equation} \label{eq:sznonoscillating}
\bar{s}_{\alpha}(x) = -\frac{1}{\sqrt{2}\pi} (\partial_x \Phi_s)
\end{equation}
and the oscillatory part by
\begin{equation} \label{eq:szoscillating}
\tilde{s}_{\eta\alpha}(x) =  \frac{1}{\pi a}
                \sin[\sqrt{2}\Phi_s]
                \sin[ 2 k_F(x-x_\eta) -\sqrt{2}\Phi_c]\;.
\end{equation}
The difference between the copper and the oxygen sites is reflected in the value of
$x_\eta$ and affects the oscillatory part; indeed, for copper $x_a=x_A=a/2$ and
for oxygen $x_b=x_B=0$ as a consequence of the different phase factors in
(\ref{eq:coefproj}).

In (\ref{eq:bosquad}), the $u_\nu$ are the new velocities for the $\nu$-excitation and the $K_\nu$ are the Luttinger liquid
parameters controlling the anomalous exponents in the correlation
functions. For weak coupling, they are related to the interactions in (\ref{eq:hamlowU})
by
\begin{eqnarray} \label{eq:lutparam}
u_s K_s & = & u_c K_c = v_F \nonumber \\
u_s/ K_s & = & v_F - a  U_o / \pi \nonumber\\
u_c/ K_c & = &  v_F + a U_o / \pi \;.
\end{eqnarray}
Since the Luttinger liquid representation is more general than the
perturbative result for small interactions, it is also applicable when
the interactions are strong. The quadratic Hamiltonian can be viewed in
this case as an effective Hamiltonian describing the low-energy properties
of the system, provided that the correct Luttinger liquid parameters are
used. Such a smooth connection between weak and strong coupling has been
proven for single-band models, \cite{haldane_xxzchain,schulz_hubbard_exact}
and a similar Luttinger representation has been shown to work for the case
of the two-band model. \cite{stechel_CuO_model,sudbo_cuo}
Equations (\ref{eq:boscomp}--\ref{eq:bosintN}) define the four-band model, 
and the NMR properties can be computed through $H_N$.

\subsubsection{Correlation functions at zero temperature}

We focus here on the spin-spin correlation functions relevant for NMR
and for neutron scattering experiments. The general form of these functions
is
\begin{equation} \label{eq:spincorrel}
R_{\eta\eta^\prime}(x,\tau )= \left\langle T_\tau S_{\eta^\prime}^z(x,\tau
)S_{\eta}^z(0,0) \right\rangle- \left\langle
S_{\eta^\prime}^z \right\rangle \left\langle
S_{\eta}^z \right\rangle \;,
\end{equation}
and it describes correlations between different orbitals $\eta$ and
$\eta^\prime$ at different points in Euclidean space-time. Here we
introduce the decomposition of this function into a non-oscillatory and an oscillatory part
\begin{equation} \label{eq:composition_spincorrel}
R_{\eta\eta^\prime}(x,\tau )=\bar{R}_{\eta\eta^\prime}(x,\tau )+\cos( 2 k_F x)\tilde{R}_{\eta\eta^\prime}(x,\tau )\;,
\end{equation}
since the behavior of these functions will be very different for
one-dimensional systems. 

Because of the doping dependence in the
cosine terms in (\ref{eq:bosintc}), the behavior of the system will quite clearly be
different for zero and for finite doping.
At half-filling, one sees from
(\ref{eq:lutparam}), (\ref{eq:massive_inequality}), and
(\ref{eq:massless_inequality}) that charge excitations are massive ($c_m$),
whereas spin excitations are in the massless regime ($s_o$). One recovers the standard Mott or
charge-transfer insulator with the massless excitations
corresponding to a Heisenberg-like exchange.
In the doped case, the term (\ref{eq:bosintc}) is irrelevant
because of the oscillatory factor $\delta x$. However, at short distances or
for short times this term is still small,
and the cosine term will influence the behavior of the system.
We thus distinguish between two different regimes for the doped case: we
assume that for intermediate distances ($a \ll x \ll l_\delta$) the system
remains in the ($c_m,s_o$)-phase as mentioned before for the half-filled
case, and when distances are larger than $l_\delta$, the system will be in
the ($c_o,s_o$)-phase because the umklapp process becomes
ineffective. The characteristic length separating these two regimes denotes
essentially the distance between two charge domain walls and is given by
$l_\delta =2 \pi/\delta$.

Due to the spin-charge separation in (\ref{eq:boscomp}), each
part of the correlation function
(\ref{eq:composition_spincorrel}) will factorize into independent averages over
the spin ($s_o$) and the charge sector ($c_o$ or $c_m$)
, and will only depend on the characteristic distance
$r_\nu=[(u_\nu \tau)^2+x^2]^{\frac{1}{2}}$  between two points in Euclidean
space-time (with $\nu=s,c$). Details about
the correlation functions in the various regimes ($c_i,s_i$) are explained in Appendix~\ref{sine-gordon-model}.
Substituting (\ref{eq:spincomp}--\ref{eq:szoscillating}) in
(\ref{eq:spincorrel}), the non-oscillatory contribution to the correlation
function is given by
\begin{equation} \label{eq:bar_R_eta_eta}
\bar{R}_{\eta\eta^\prime}=|\lambda_{\eta\alpha}|^2|\lambda_{\eta^\prime\alpha}|^2\bar{R}_\alpha(r_s)\;,
\end{equation}
where $\bar{R}_\alpha(r_s)=(2\pi r_s)^{-2}$ depends only on the
spin degrees of freedom and is thus completely independent of the coexisting
charge phase. Notice that this function is also independent of the orbitals
$\eta$ and $\eta^\prime$, and thus there is no fundamental difference between
copper and oxygen contributions.

For the oscillatory part of the spin-spin correlation functions, the situation will be
quite different. We restrict ourselves to the calculation of correlation
functions between identical orbitals ($\eta=\eta^\prime$). Using averages
over the charge and spin sectors of the Hamiltonian (\ref{eq:boscomp}),
these functions can be reexpressed  as
\begin{equation}\label{eq:massive_correlation}
\tilde{R}_{\eta\eta}^{c_m,s_o} = \frac{|\lambda_{\eta\alpha}|^4}{(2 \pi a)^2}
\tilde{R}_{\eta\alpha}^{c_m}(r_c)\tilde{R}_{\alpha}^{s_o}(r_s)
\end{equation}
in the massive charge regime and as
\begin{equation}
\tilde{R}_{\eta\eta}^{c_o,s_o} = \frac{|\lambda_{\eta\alpha}|^4}{(2 \pi a)^2}
\tilde{R}_{\alpha}^{c_o}(r_c)\tilde{R}_{\alpha}^{s_o}(r_s)
\end{equation}
in the massless charge regime. The newly defined correlation functions in
the massless phases ($\nu_o$) are given by
\begin{equation}
\tilde{R}_{\alpha}^{\nu_o}(r_\nu) =  (a/r_\nu)^{K_\nu^*}F(r_\nu)\;.
\end{equation}
The function $F(r_\nu)$ describes the corrections to the Luttinger
Liquid behavior which come from the flow to the fixed point
\cite{giamarchi_logs}. To lowest order, $F(r_\nu)$ can be approximated by $1$. The renormalized Luttinger liquid
parameters $K_\nu^*$ for a spin symmetric model with repulsive interaction are
restricted to
\begin{equation}
K_s^*=1 \quad{\rm and}\quad 0 \le K_c^* \le 1\;.
\end{equation}
The value of the renormalized Luttinger liquid parameter $K_c^*$
depends on the interactions. For weak interaction, $K_c^*$ is close to $1$,
and it decreases as interactions become more repulsive.

The correlation functions in (\ref{eq:massive_correlation}) which are characterized by the massive charge phase
are given by
\begin{eqnarray}
\tilde{R}_{A\alpha}^{c_m}(r_c) & = &  2 \cosh[ K_c K_0(m_c r_c)](m_ca)^{K_c}
\nonumber
\\
\tilde{R}_{B\alpha}^{c_m}(r_c) & = &  2 \sinh[ K_c K_0(m_c r_c)] (m_c
a)^{K_c}
\label{eq:oscillating_charge}
\end{eqnarray}
and depend on the chosen orbital $\eta$. Thus, the behavior for copper and
oxygen will be quite different. It depends on the distance $r_c$, the mass
$m_c$, and the stiffness constant $K_c$. In general, for distances larger
than $l_{m_c}=1/m_c$, the function $\tilde{R}^{c_m}_{A\alpha}$ for copper tends to a
finite constant, whereas $\tilde{R}^{c_m}_{B\alpha}$ for oxygen tends
exponentially to zero.

\subsubsection{The asymptotic expressions at finite temperature}

In order to obtain the temperature dependent correlation function 
$\tilde{R}_{\eta\eta}(x,\tau,\beta)$, 
we will only use the asymptotic expressions of
(\ref{eq:oscillating_charge}). We recover a Luttinger liquid behavior,
and the temperature dependence can easily be obtained with the help of the
conformal symmetry;\cite{tsvelik_quantum_field_theory} indeed, we only need to replace
$r_\nu(x,\tau)$ by $r_\nu(x,\tau,\beta)$ where
\begin{equation} \label{eq:r_nu_to_r_nu_beta}
r_\nu(x,\tau,\beta)=\frac{u_\nu \beta}{\pi}\sqrt{
\sinh\left[ \frac{x- i u_\nu \tau}{u_\nu \beta/\pi}\right] \sinh\left[
\frac{x+i u_\nu \tau}{u_\nu \beta/\pi}\right]}\;.
\end{equation}
The relevant asymptotic expressions at half-filling and away
from half-filling depend on the relative magnitudes of the various characteristic lengths of the
system, namely the lengths related to the mass, $l_{m_c}$, and to the doping,
$l_\delta$, as well as the thermal length $l_\beta={\rm min}\{ (u_c \beta)^{-1},(u_s \beta )^{-1}\}$.
For half-filling ($\gamma=0$) at low temperature, we are in the regime where $l_\beta \gg
r_c \gg l_{m_c}$ and $l_\delta=\infty$, thus we can approximate the
oscillatory charge contribution in (\ref{eq:massive_correlation}), and the
functions are simplified to
\begin{equation}\label{eq:tilde_R_eta_eta_0}
\tilde{R}_{\eta\eta}^0 =C^0_\eta |\lambda_{\eta\alpha}|^4 \tilde{R}^0_\alpha(r_s)\;,
\end{equation}
where the amplitude of the oscillatory part at half-filling is given by
$C_\eta^0=\tilde{R}_{\eta\alpha}^{c_m}(\infty_c)$. The remaining correlation
function is independent of the orbital $\eta$ and given by $\tilde{R}_\alpha^0=(2\pi a)^{-2} (a/r_s)$.

\begin{figure}
\centerline{\epsfig{file=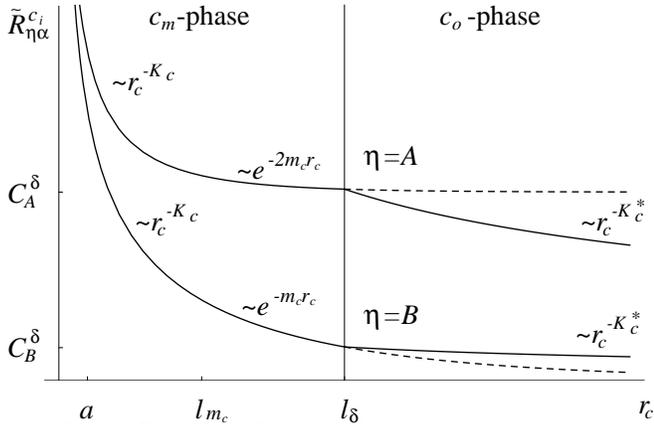,angle=0,width=8.6cm}}
\caption{Crossover for the charge contribution to
the spin-spin correlation function $\tilde{R}_{\eta\alpha}^{c_i}$ from the
massive charge regime ($c_m$-phase) to the massless charge regime
($c_o$-phase) at low temperature and for small doping. The behavior for
copper (A) and oxygen (B) is quite different: For copper the amplitude remains
finite for long distance, whereas the amplitude for oxygen vanishes.}
\label{fig:correlations}
\end{figure}

The large distance limit of the corresponding expression for small doping
($\gamma=\delta$) and low temperature, where $l_\beta \gg r_c \gg l_\delta \gg l_{m_c}$,  looks like
\begin{equation}\label{eq:tilde_R_eta_eta_delta}
\tilde{R}_{\eta\eta}^{\delta} = C_\eta^\delta |\lambda_{\eta\alpha}|^4
\tilde{R}_\alpha^\delta(r_c,r_s)\;.
\end{equation}
In the doped regime, $C_{\eta}^\delta$ is the amplitude 
$\tilde{R}_{\eta\alpha}^{c_m}(l_\delta)$ obtained in the massive
phase at the crossover, as shown in Fig.~\ref{fig:correlations}. Like
before, the remaining correlation function
$\tilde{R}_\alpha^\delta=(2\pi a)^{-2}(a/r_s)(a/r_c)^{K^*_c}$ is also
independent of $\eta$ but shows dependence on spin and charge degrees of
freedom. For larger doping rates ($l_\delta < l_{m_c}$), the difference between
copper and oxygen sites vanishes.

\subsubsection{Knight shifts and relaxation rates}

The standard expressions for the Knight shifts and for the relaxation rates
resulting from a hyperfine coupling term like (\ref{eq:hamN}) are
\begin{eqnarray} \label{eq:knightshift}
K_\eta^\gamma & = & \frac{C_\eta}{\gamma_\eta \gamma_e \hbar^2}
\sum_{\eta^\prime=a,b} \chi_{\eta\eta^\prime}^\gamma(\omega=0,p \to 0)
\\
\label{eq:relaxationtime}
\frac{1}{T_{1\eta}^\gamma} & = & \frac{C_\eta^2}{\gamma_\eta \gamma_e \hbar^2 \beta} \sum_{p}
\frac{{\rm Im}\left[ \chi_{\eta\eta}^\gamma(\omega_\eta,p)\right]}{\omega_\eta}
\;,
\end{eqnarray}
where $\gamma=0$ refers to the half-filled case and $\gamma=\delta$ to the
doped case. $\omega_\eta$ denotes the electronic Zeeman frequency in orbital
$\eta$, which is
very small as compared to the energy scale of the purely electronic system
fixed by the cutoff $\lambda$. For the Knight shifts the sum is restricted to the active
orbitals $a$ and $b$.
We can split up the susceptibility
$\chi_{\eta\eta^\prime}^\gamma$ into the non-oscillatory
$\bar{\chi}_{\eta\eta^\prime}^\gamma$ and the oscillatory contribution
$\tilde{\chi}_{\eta\eta^\prime}^\gamma$ just like for the correlation functions in (\ref{eq:composition_spincorrel}).
Finally, the Knight shifts for the linearized four-band model in units of $C_\eta/(\gamma_\eta \gamma_e \hbar^2)$
are given by
\begin{equation}
\label{eq:K_eta_gamma}
K_\eta^\gamma = \bar{F}_{\eta\alpha}\bar{\chi}_{\alpha}(\omega=0,q \to0)
\end{equation} and the relaxation rates in units of $C_\eta^2/(\gamma_\eta
\gamma_e \hbar^2 \omega_\eta)$ by
\begin{equation}
\label{eq:T_1_eta_gamma}
\frac{1}{T_{1\eta}^\gamma} =  \frac{1}{\beta} \sum_{|q|<\lambda}
{\rm Im}\left[ (\bar{F}_{\eta\alpha})^2 \bar{\chi}_\alpha(\omega_\eta,q)+
(\tilde{F}_{\eta\alpha}^\gamma)^2 \tilde{\chi}_\alpha^\gamma(\omega_\eta,q) \right]\;.
\end{equation}
The susceptibilities $\bar{\chi}_\alpha$ and $\tilde{\chi}_\alpha^\gamma$ in
space-time can be obtained from
\begin{eqnarray}\label{eq:bar_chi_alpha}
\bar{\chi}_\alpha(x,t)=2\theta(t){\rm
Im}\big[\bar{R}_\alpha(x,\tau,\beta)\big]_{\tau=it+\epsilon}
\\      \label{eq:tilde_chi_alpha}
\tilde{\chi}_\alpha^\gamma(x,t)=
2\theta(t){\rm
Im}\big[\tilde{R}_\alpha^\gamma(x,\tau,\beta)\big]_{\tau=it+\epsilon}
\end{eqnarray}
performing the continuation to real time.
The $\tau$-ordered temperature dependent Green's functions on the right-hand
side are the same as in
(\ref{eq:bar_R_eta_eta}), (\ref{eq:tilde_R_eta_eta_0}) and
(\ref{eq:tilde_R_eta_eta_delta}), using 
(\ref{eq:r_nu_to_r_nu_beta}). Thus, in general we can calculate (\ref{eq:K_eta_gamma}) and (\ref{eq:T_1_eta_gamma}) by performing the Fourier transform of (\ref{eq:bar_chi_alpha}) and
(\ref{eq:tilde_chi_alpha}).
Here we restrict ourselves to the solutions obtained by the so called power
counting method. The temperature dependences of the
Knight shifts and of the relaxation rates are shown in
Table~\ref{table:knightshifts_relaxationrates_small_U} and the form
factors $\bar{F}_{\eta\alpha}$ and $\tilde{F}_{\eta\alpha}^\gamma$ are given
in Table~\ref{table:formfactors_general_small_U}.

\subsection{NMR properties of the Bulut model}
\label{NMR_properties_Bulut}
In order to obtain the NMR properties of the 1D version of the Bulut model
 we perform the same
procedure as before for the four-band model. The bosonized version of the
Bulut model (\ref{eq:hamcompBu}) is given by (\ref{eq:boscomp}) replacing
$H_N$ by
\begin{equation} \label{eq:h_N_bulut_bosonized}
H_N^{Bu} =\int dx \left[ C_A{\bf I}(x){\bf S}_{A}^{Bu}(x)+C_B{\bf J}(x){\bf S}_{B}^{Bu}(x) \right]\;,
\end{equation}
where ${\bf S}_{\eta}^{Bu}$ is the projection onto the $\alpha$-band
using (\ref{copper_spin_approx_bulut}), (\ref{eq:oxygen_spin_approx_bulut}),
(\ref{eq:spindensity}) and (\ref{eq:proj}). For the z-component of the
spin ${\bf S}_{\eta}^{Bu}$ we get
\begin{eqnarray}
S_{A}^{zBu}(x) &=& 2
|\lambda_{Aa}^{Bu}|^2|\lambda_{a\alpha}|^2[\bar{s}_\alpha(x)+\cos(
2k_Fa)\tilde{s}_{a\alpha}(x)]
\nonumber
\\
S_{B}^{zBu}(x) &=& 2 |\lambda_{Ba}^{Bu}|^2|\lambda_{a\alpha}|^2[\bar{s}_\alpha(x)+\cos( k_Fa)\tilde{s}_{a\alpha}(x)]
\nonumber
\\
\end{eqnarray}
ignoring all gradient terms of the field $\phi_c$. The spin operators
$\bar{s}_\alpha$ and $ \tilde{s}_{\eta\alpha}$ are defined as before in
(\ref{eq:sznonoscillating}) and (\ref{eq:szoscillating}).
The NMR properties for a hyperfine coupling like (\ref{eq:h_N_bulut_bosonized}) are
given by (\ref{eq:K_eta_gamma}) and (\ref{eq:T_1_eta_gamma}) by the replacements
$\bar{F}_{\eta\alpha} \to \bar{F}_{\eta\alpha}^{Bu}$ and
$\tilde{F}_{\eta\alpha}^\gamma\to \tilde{F}_{\eta\alpha}^{\gamma Bu}$. The values for the different
form factors are shown in Table~\ref{table:formfactors_bulut}. At this level
of approximation, both the four-band and the Bulut model show exactly the same
temperature dependence for the Knight
shifts as well as for the relaxation rates; this dependence is different for
the uniform contribution and for the oscillatory one (see
Table~\ref{table:knightshifts_relaxationrates_small_U}), as is well known
for interacting one-dimensional systems. This effect has nothing to do with
the various orbitals where the Knight
shifts and the relaxation rates are measured.

\subsection{Comparing the four-band model and the 1D Bulut model}
\label{general_Bulut}
First we focus on the coefficients of the
Bulut and of the four-band model (compare
Table~\ref{table:formfactors_general_small_U} and
\ref{table:formfactors_bulut}) related to the different projection
procedures of the s-orbitals onto the ground state. For comparing both
models, we investigate the limit
$t_{ab}\ll(\epsilon_b-\epsilon_a)$. Then, for the four-band model the
projection of the s orbitals ($A,B$) onto the lowest band ($\alpha$) is
strictly done in $k$-space and results in
\begin{eqnarray}
|\lambda_{B\alpha}(k_F)|^2 &\to &
 2      (\lambda_{Ba}^{Sh})^2
\\
&-&2  \cos(k_Fa) (\lambda_{Ba}^{Sh})^2
\label{eq:O_general_dynamic}
\\
\nonumber\\
|\lambda_{A\alpha}(k_F)|^2 & \to &
2       (\lambda_{Aa}^{Sh}+\lambda_{Aa}^{Mi})^2
\\
&+&4    (\lambda_{Aa}^{Mi})^2
\label{eq:Cu_general_local_additional}
\\
&+&2  \cos(2k_Fa)   (\lambda_{Aa}^{Sh}+\lambda_{Aa}^{Mi})^2
\label{eq:Cu_general_dynamic}
\\
&+&8   \cos(k_Fa)   (\lambda_{Aa}^{Sh}+\lambda_{Aa}^{Mi})\lambda_{Aa}^{Mi}
\label{eq:Cu_general_dynamic_combined}\;,
\end{eqnarray}
whereas for the Bulut model it is a combination of real space and
$k$-space projection yielding
\begin{eqnarray}
2|\lambda_{Ba}^{Bu}|^2|\lambda_{a\alpha}(k_F)|^2 &\to& 2
(\lambda_{Ba}^{Sh})^2
\label{eq:O_bulut_transferred}
\\
\nonumber\\
2|\lambda_{Aa}^{Bu}|^2|\lambda_{a\alpha}(k_F)|^2
&\to& 2(\lambda_{Aa}^{Sh}+\lambda_{Aa}^{Mi})^2 \label{eq:Cu_bulut_transferred}\;.
\end{eqnarray}
The general solution for the projected O-3s orbital includes one
more term (\ref{eq:O_general_dynamic}) than the solution proposed by Bulut (\ref{eq:O_bulut_transferred}). This term
corresponds to a dynamic contribution which includes a charge
displacement. However, for a half-filled system the additional term vanishes
and the two solutions become identical. By contrast, the projection
procedure for the Cu-4s orbital produces a completely different behavior in
the two models. For a half-filled system, the hopping processes via
$\lambda_{Aa}^{Bu}=\lambda_{Aa}^{Sh}+\lambda_{Aa}^{Mi}$ contribute only in
the Bulut model (\ref{eq:Cu_bulut_transferred}), whereas they are exactly canceled by the related dynamic terms
(\ref{eq:Cu_general_dynamic}) in the four-band model. Thus, for the four-band model at half-filling,
only an additional local term (\ref{eq:Cu_general_local_additional}) as
well as a dynamic combined term (\ref{eq:Cu_general_dynamic_combined})
remain.
The term (\ref{eq:Cu_general_local_additional}) is the local analog to the
transferred terms proposed by Mila-Rice, and the term
(\ref{eq:Cu_general_dynamic_combined}) is a combination of Mila-Rice and
Shastry terms which includes a charge displacement.
It should be clear that our projection procedure is the right one for a
system with small Coulomb interactions: First we diagonalize the
tight-binding Hamiltonian dealing with extended wave functions, and then we
treat the Coulomb energy approximately within this non-local basis. The
approximation proposed by Bulut suffers from a mismatch between the local
and the non-local point of view.

The second part of the oscillatory contribution to the form factors  (compare
Table~\ref{table:formfactors_general_small_U} and
\ref{table:formfactors_bulut}), which contains the dependence on the
characteristic lengths related to the doping rate, $l_\delta$, as well as to
the charge mass, $l_{m_c}$, is the crucial one.
Away from half-filling, the four-band model shows a different behavior on
the copper and on the oxygen, despite the fact that umklapp processes
only contribute
on short or intermediate scales. Indeed, the
different hyperbolic dependencies of the two characteristic lengths
$l_{m_c}$ and $l_\delta$ for copper and for
oxygen (see Table~\ref{table:formfactors_general_small_U})
affect measured quantities related to long distance
or long time behavior. Instead, for the Bulut model the difference between copper and oxygen comes
in only because of the special choice of a Mila-Rice-Shastry type
electron-nuclear interaction term (\ref{eq:h_N_bulut_bosonized}) and the related
unconventional projection procedure which results in the different trigonometric form
factors (see Table~\ref{table:formfactors_bulut}). The influence of the charge
mass $m_c$ is the same for copper and for oxygen, a fact  which manifests 
itself by the same dependence on the characteristic length $l_{m_c}$.

Note that the four-band model leads to a very small contribution on the
oxygen even at finite doping, because the
contribution is exponentially suppressed in a way which depends on the ratio between
$l_{m_c}$ and $l_\delta$, whereas the oscillatory contribution on the copper
atom is nearly independent of the doping rate for long distances or
times. For the ratio between copper and oxygen we distinguish between two regimes:
\begin{equation}
\tanh\left[K_c K_0\left( \frac{l_\delta}{l_{m_c}}\right)\right] \to
\left\{
\begin{array}{l@{\quad {\rm for} \quad}l}
1 & l_\delta \ll l_{m_c}
\\
0 & l_\delta \gg l_{m_c}
\end{array}
\right. \;.
\end{equation}
In the former regime, we recover the Luttinger Liquid behavior, since the
infinite length $l_{m_c}$ stems from the vanishing of the umklapp process
when  $U_{a\alpha}=U_{b\alpha}$; in that case there
is no fundamental difference between
copper and oxygen anymore. Only the overlaps with the ground state remain
different. The latter regime, where the fundamental difference occurs, will
be reached exponentially as $K_c \sqrt{\pi l_{m_c}/ 2 l_\delta}\exp(-l_\delta/l_{m_c})$, and thus the oxygen does not see the
antiferromagnetic fluctuations in this limit. Instead,
 for the Bulut model everything depends on the same
correlation function, and the difference between copper and oxygen comes
from the filtering factors. Thus, the oscillating
contributions to the relaxation rates for the oxygen is always
proportional to $[0+(\pi a/ 2 l_\delta)^2]$, whereas the contributions on copper
are reduced by a factor $[1-(\pi a/ l_\delta)^2]$. The ratio of
the oscillating contribution to the relaxation rates is approximately given
by $(\pi a/2l_\delta)^2$, and is completely independent of the details of
the projected local Coulomb repulsions $U_{a\alpha}$ and $U_{b\alpha}$.
It only depends on the doping rate and is proportional to
$(\delta a /4)^2$. By contrast, the four-band
model includes the effect of the Coulomb interactions through its
dependence on $l_{m_c}$. In Table~\ref{table:ratios} we show the ratios of the different Knight shifts and relaxation
rates contributions.
\section{The strong interaction limit}\label{strong_interaction_limit}

\begin{figure}
\centerline{\epsfig{file=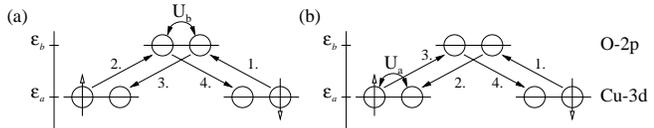,angle=0,width=8.6cm}}
\caption{Superexchange processes generating  antiferromagnetic
couplings between localized copper spins at half-filling. In contrast to
the superexchange path (a)  process (b) is
suppressed, because it includes an intermediate state where it is necessary
to pay the local Coulomb repulsion $U_a$. The numbers (1.,2.,3., and
4.) denote the sequence of the intermediate steps.}
\label{fig:process_superexchange}
\end{figure}

For strong interactions the four-band system in (\ref{eq:hamcomp}) is much more
difficult to solve.
Yet, it is still possible to highlight the qualitative features of
 the transferred
hyperfine coupling interaction, specifically for the half-filled case. To obtain the strong interaction limit of
this model we can perform the Gutzwiller projection
\begin{equation} \label{eq:projected_H}
H=\hat{P} (H_0 + H_S + H_N )\hat{P}
\end{equation}
which eliminates doubly occupied states in the Cu-3d orbitals
from the Fock space. The projection is effectively performed on all
three terms of (\ref{eq:projected_H}), which are treated on equal footing.
As far as the first part $\hat{P} H_0 \hat{P}$ is concerned, two possible
superexchange processes are generated, as shown in
Fig.~\ref{fig:process_superexchange}. In the
strong interaction limit ($U_a \gg
|\epsilon_\eta-\epsilon_{\eta^\prime}|, U_b\gg t_{\eta\eta^\prime}$) the
superexchange process in Fig.~\ref{fig:process_superexchange}(a) is much more
effective than the process in Fig.~\ref{fig:process_superexchange}(b).
For the basic system $H_0$, we only have to keep 3 states per unit cell,
whereas for the four-band model (\ref{eq:projected_H}), 
we end up with a system where we have to keep 27 spin-degenerate local
states per unit cell $j$ with 4 tight-binding parameters
$t_{\eta\eta\prime}$ for a half-filled system (excluding doubly excited
$A,B$-states, see Appendix~\ref{local_states_for_u_a}).
For a doped system the number of states as well as the number of possible
transitions increases very fast, as has been shown for
a two-band model.\cite{zaanen_canonical_pertubation_for_highTc}
A correct projection procedure such as (\ref{eq:projected_H})
becomes very difficult to handle, and one must resort to some approximations.
In any event, in the vicinity of the  half-filled case where the projection
can be explicitly used for the full Hamiltonian, we will analyze the
differences between the predictions of the four-band model and those of the
approximated Hamiltonians.
So let us restrict our analysis to the half-filled case where
only virtual double occupancies of the copper site are allowed and where 
electron-nuclear interaction processes require that the initial and
the final charge distribution be the same. We deal with electron-nuclear
interaction processes where effectively one local Cu-3d spin will be
reversed and then relaxed by the thermodynamic fluctuations of the Heisenberg model. We decompose $(H_0+H_S+H_N)$ into $(L+K)$.
$L$ includes all local and $K$ all kinetic
contributions of the complete Hamiltonian $H$ introduced in
(\ref{eq:hamcomp}). Then we can expand  $\hat{P}H\hat{P}$
on the basis of the unperturbed eigenstates of $L$ and compute
 the projected local s-orbital spin operators like $\hat{P}{\bf S}{\eta
j}\hat{P}$. For the details we refer to Appendix~\ref{local_states_for_u_a} and discuss only the final results.

First we analyze some relaxation processes for the oxygen atom. The process
shown in {Fig.~\ref{fig:process_oxygen}(a)} is a
transferred (T) contribution proportional to
\begin{equation}
F_{B,T,(a)}=\left[ \frac{t_{Ba}}{\epsilon_B-(\epsilon_a+U_a)}\right]^2\;.
\end{equation}
For the process shown in {Fig.~\ref{fig:process_oxygen}(b)}, we include a part of the superexchange process to avoid
double occupation of the copper site, and the contribution is
proportional to
\begin{equation}
F_{B,T,(b)}=\left[ \frac{t_{ab}t_{Ba}}{(\epsilon_a-\epsilon_b)(\epsilon_B-\epsilon_b)}\right]^2
\;.
\end{equation}
Then the lowest order contribution to the general form factor for the
oxygen is given by
\begin{equation} \label{eq:formfactor_oxygen_torsten}
F_B(p) = 2 F_{B,L} + 2 F_{B,T} \cos(pa/2)
\end{equation}
with
\begin{eqnarray}
F_{B,L} &=& 0
\\
\label{F_BT}
F_{B,T} &=& \underbrace{n_{B,T,(a)} F_{B,T,(a)}}_{\rm projected\; Shastry}+
n_{B,T,(b)} F_{B,T,(b)}+  \ldots \;.
\end{eqnarray}
$n_{B,T,(i)}$ denotes the combinatorial factor 
associated with all possible processes yielding a
contribution $ F_{B,T,(i)}$. The factor 2 for the left-right
symmetry is not included in $n_{B,T,(i)}$. Like for the
superexchange processes (Fig.~\ref{fig:process_superexchange}) some processes are forbidden due to the Pauli principle. However, since all
energy levels are assumed to be spin-independent the related amplitudes
$F_{B,T,(i)}$ are the same.

\begin{figure}
\widetext
\centerline{\epsfig{file=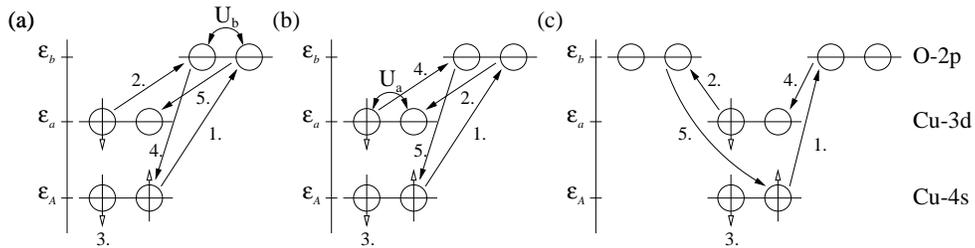,angle=0,width=12.9cm}}
\caption{Possible local spin flip processes of the nuclear copper spin
like $\hat{P}S^+_{Aj}\hat{P}$ via a Fermi
contact interaction at half-filling. All processes include the O-2p orbital
as an intermediate state and are of the same order as the transferred
hyperfine coupling processes proposed by Mila and Rice.}
\label{fig:process_copper_local}
\end{figure}

For copper we also distinguish between the transferred
(Fig.~\ref{fig:process_copper_trans}) and the local contributions
(Fig.~\ref{fig:process_copper_local}). The transferred contributions are
proportional to
\begin{eqnarray}
F_{A,T,(a)} & = & \left[
\frac{t_{Ab}t_{ab}}{(\epsilon_A-\epsilon_b)(\epsilon_A+\epsilon_a-2\epsilon_b-U_b)}
\right]^2 \label{eq:process_AT(a)}
\\
F_{A,T,(b)} & = & \left[
\frac{t_{Ab}t_{ab}}{(\epsilon_A-\epsilon_b)(\epsilon_A-\epsilon_a-U_a)}
\right]^2 \label{eq:process_AT(b)}
\\
F_{A,T,(c)} & = & \left[
\frac{t_{Aa}}{\epsilon_A-\epsilon_a-U_a}
\right]^2 \label{eq:process_AT(c)}
\\
F_{A,T,(d)} & = &
\frac{t_{Ab}t_{ab}t_{Aa}}{(\epsilon_A-\epsilon_b)^2(\epsilon_a-\epsilon_b)}
\;,\label{eq:process_AT(d)}
\end{eqnarray}
whereas the local contributions are given by
\begin{eqnarray}
F_{A,L,(a)} & = & \left[
\frac{t_{Ab}t_{ab}}{(\epsilon_A-\epsilon_b)(\epsilon_A+\epsilon_a-2\epsilon_b-U_b)}
\right]^2 \label{eq:process_AL(a)}
\\
F_{A,L,(b)} & = & \left[
\frac{t_{Ab}t_{ab}}{(\epsilon_A-\epsilon_b)(\epsilon_A-\epsilon_a-U_a)}
\right]^2 \label{eq:process_AL(b)}
\\
F_{A,L,(c)} & = & \left[
\frac{t_{Ab}t_{ab}}{(\epsilon_A-\epsilon_b)(\epsilon_A+\epsilon_a-2\epsilon_b)}
\right]^2\;.\label{eq:process_AL(c)}
\end{eqnarray}
Then, the general form factor for copper reads
\begin{equation} \label{eq:formfactor_copper_torsten}
F_A(p) = 2 F_{A,L} + 2 F_{A,T} \cos(pa)
\end{equation}
with
\begin{eqnarray}
F_{A,L} &=& n_{A,L,(a)}F_{A,L,(a)}+ n_{A,L,(b)}F_{A,L,(b)}
\nonumber\\
&+&n_{A,L,(c)}F_{A,L,(c)}+\ldots
\label{F_AL}
\\ \nonumber
\\ \nonumber
F_{A,T} &=& \underbrace{n_{A,T,(a)}F_{A,T,(a)}+ n_{A,T,(b)}F_{A,T,(b)}}_{\rm
projected\; Mila-Rice}
\nonumber\\
&+&\underbrace{n_{A,T,(c)}F_{A,T,(c)}}_{\rm projected\; Shastry}+ n_{A,T,(d)}F_{A,T,(d)}+\ldots
\;.
\label{F_AT}
\end{eqnarray}
Let us now compare the predictions of the four-band model and those of the
 1D Mila-Rice or Shastry models\label{comparison_g_MRS}.

Using the projected expression for the oxygen  
instead of the unprojected one (\ref{eq:coef_shastry_Ba}), 
only process (b) in (\ref{F_BT})
contributes in the strong interaction limit, whereas process (a)  in
(\ref{F_BT}) proposed by Shastry becomes negligible
\begin{equation}
F_{B,T,(a)}\stackrel{U_a\to \infty}{\longrightarrow}0\;.
\end{equation}
The form factor for the characteristic wave vectors ($p=0$ or $p=\pi/a$)
is then reduced to
\begin{eqnarray}
F_B(0) &=& 2n_{B,T,(b)}F_{B,T,(b)}\\
F_B(\pi/a) &=& 0\;.
\end{eqnarray}
Since only the relaxation process of the oxygen nuclear
 spin contributes, which corresponds to $p\sim0$ , we recover
the basic structure of the form factor of Shastry with modified
amplitudes. Thus at half-filling, there is no fundamental
difference for the oxygen between the general form
factor (\ref{eq:formfactor_oxygen_torsten}) and the form factor proposed by
Shastry (\ref{eq:formfactor_oxygen_shastry}).

In the strong coupling limit at half filling, the following contributions to
the form factor for copper are suppressed:
\begin{equation}
F_{A,L,(b)}, F_{A,T,(b)}, F_{A,T,(c)} \stackrel{U_a\to \infty}{\longrightarrow} 0 \;.
\end{equation}
Thus the projected Shastry contribution (c) in (\ref{F_AT}) and one of the
projected Mila-Rice contributions (b) in (\ref{F_AT}) as well as one of the projected local
contributions (b) in (\ref{F_AL}) become negligible, and we end up with
\begin{eqnarray}
F_{A,L} &=& n_{A,L,(a)}F_{A,L,(a)}+n_{A,L,(c)}F_{A,L,(c)}+\ldots
\\
F_{A,T} &=& n_{A,T,(a)}F_{A,T,(a)}+n_{A,T,(d)}F_{A,T,(d)}+\ldots
\end{eqnarray}
for the local and for the transferred contributions
 to the general form factor
(\ref{eq:formfactor_copper_torsten}), respectively. Thus the uniform part of the form
factor is given by
\begin{eqnarray}
F_A(0) &=& 4 n_{A,(a)}F_{A,(a)}+2 n_{A,L,(c)}F_{A,L,(c)}
\nonumber\\
&+& 2 n_{A,T(d)}F_{A,T(d)}+\ldots\;,
\end{eqnarray}
whereas the oscillating part reads
\begin{equation} \label{eq:F_A_antiferro}
F_A(\pi/a) = 2 n_{A,L,(c)}F_{A,L,(c)} - 2 n_{A,T,(d)}F_{A,T,(d)}+\ldots\;.
\end{equation}
We used the fact that $n_{A,L,(a)}=n_{A,T,(a)}\equiv n_{A,(a)}$ and
$F_{A,L,(a)}=F_{A,T,(a)}\equiv F_{A,(a)}$.
The uniform part includes contributions
which are absent in the 1D version of the Mila-Rice and of the Shastry
model. Furthermore, some terms proposed by Shastry turn out to be zero
in the strongly interacting limit.
For the oscillatory part the effects are much more drastic. The
transferred terms proposed by Shastry vanish in the strong coupling regime,
whereas other transferred terms, which come from a combination of Mila-Rice
and Shastry processes, contribute. Besides, the transferred terms proposed
by Mila and Rice are canceled by the equivalent local terms. Hence, in 1D, the
general form factor differs both qualitatively and quantitatively 
from the form factors one would derive from the Mila-Rice or from the Shastry
models.

\section{Discussion and perspectives}\label{discussion and perspectives}

In this paper, we have analyzed the 1D analogs of the hyperfine form factors
 proposed
for NMR measurements of high-$\rm{ T_c}$ materials in the antiferromagnetic
phase. We have focused on the situation where
one deals with an antiferromagnet generated by a superexchange process via
 an oxygen atom 
located at the midpoint between two copper atoms and where the Fermi contact
interaction is one of the main contributions to the possible electron-nuclear
interaction terms. We have
investigated a 1D Cu-O model including four orbitals per
unit cell, namely the Cu-3d and the O-2p orbitals governing the ground state
properties, as well as the Cu-4s and O-3s orbitals describing the isotropic
Fermi contact interaction. In 1D, we were able to solve this model using only
standard techniques without having to introduce any additional approximations for the hyperfine
interaction term as proposed by Mila-Rice and by Shastry. Thus, we were able to
compare our solutions of the four-band model with the predictions of the approximative models.

In the low interaction limit, we have calculated the resulting
temperature dependence of the Knight shifts $K$ and of the 
relaxation rates $1/T_1$ for an undoped and for a doped
system; in that limit the ground state is
well described by the strongly hybridized Cu-3d--O-2p anti-bonding
band the width of which is large as compared to all Coulomb interactions.
For both models, the four-band and the approximative one (Bulut model), the temperature
dependences are the same and show the typical power law behavior of one-dimensional interacting systems (see Table~\ref{table:knightshifts_relaxationrates_small_U}).
Within this scope we have shown for the four-band model that for an undoped and a
 slightly doped system copper and oxygen behave completely different for
 long distances or long times, when the temperature is low enough.
The oxygen nuclei see only the Korringa-like contributions, since the
 antiferromagnetic contributions are exponentially suppressed depending on
the ratio of the characteristic length related to the charge gap and the
doping. In contrast, the copper nuclei always see both contributions, the
 Korringa-like contribution as well as the antiferromagnetic one. This
 fundamental difference between copper and oxygen vanishes gradually when
 the characteristic doping length or the characteristic thermal length
 becomes shorter than the length related to the charge gap (the difference 
 goes away abruptly when the system develops a gap in the spin sector). 
This solution is at variance with the prediction of the related
 approximate model,
where  for oxygen the antiferromagnetic contributions to $1/T_1$ increase
with doping like $\delta^2$, whereas  for copper 
they decrease proportionally
to $\delta^2$. Thus, the scenario where oxygen does not see the
antiferromagnetic fluctuations is realized much more effectively in the
four-band model than in the 1D version of the models proposed for the
high-$T_{\rm c}$ materials. In 1D, such an unconventional
scenario works, since even small interactions generate
strong antiferromagnetic correlations due to the drastic reduction of
the Fermi surface.

We have also considered the strong interaction limit. 
Performing a Gutzwiller projection onto
the four-band model without further approximations for the electron-nuclear
interaction term, we computed the various processes contributing to
NMR. Our analysis was limited to 
the insulating phase (Heisenberg model), since even in 1D a full solution
of the model for a doped system (t-J model with four orbitals per unit cell)
is unavailable.
In the strong interaction limit of the 1D Cu-O model, we were able to compare
the form factors obtained for the four-band model with the predictions
obtained for the approximate models (Mila-Rice model and Shastry model)
investigating the different relaxation processes for the copper and
oxygen nuclear spins. In this context, we have shown that neither the 1D analog
of the Mila-Rice model nor the 1D analog of the Shastry model could describe
the strong interaction limit at half-filling. In contrast to the usual
assumption that only transferred contributions are relevant, we
predict that both local and transferred contributions should be taken into
account for describing the relaxation of the nuclear copper spin via an Cu-4s
orbital. Furthermore, we have shown that for infinite local repulsions on the
copper sites and small local repulsions on the oxygen sites, the
contributions proposed by Mila-Rice and Shastry vanish. For the relaxation of
the nuclear oxygen spin we recover the
basic idea of transferred hyperfine couplings with slightly modified
amplitudes, but once again the contribution proposed by
Shastry vanishes for infinite repulsion on the copper site.

Both the strong and the weak coupling limits underscore the importance of
keeping the full four-band model, at least in one dimension,
 in order to give an
accurate description of the NMR properties. The method we used in the present
paper to tackle such a model
can thus be extended in various
directions. First, it can be applied to study specific models which have
a structure similar to the model Cu-O chain analyzed here. This is for
example the case for ladder materials such as ${\rm Sr_{14-x}Ca_x
Cu_{24}O_{41}}$. Analyses of the NMR material have so far been
performed in terms of Mila-Rice-Shastry approximations.
An analysis retaining the full four-band model, with the specific
symmetries of these ladder systems, is currently in progress. \cite{unpublished_becker_1}
Other systems for which our analysis can be relevant are TMTSF and
TMTTF alloys. \cite{jerome_revue_1d} At stoichiometric composition they form
an alternate stack.\cite{Ilakovac_TMTSF_TMTTF} 
Let us now comment on anisotropic contributions to $K$ and to $1/T_1$; these can be
produced by a dipolar hyperfine coupling (see Appendix~\ref{anistropic_hyperfin}). They also stem from the
specific structural details of a given compound which may lead to an 
anisotropic form for the susceptibility: in that situation the anisotropy 
of the $p=0$ component (\ref{eq:bar_chi_alpha})
will usually be different from that of the $p=2k_F$ part (\ref{eq:tilde_chi_alpha}). In both
the weak and the strong interaction limits, we find that -- for 
low enough temperature -- $1/T_1$ is
mostly determined by (\ref{eq:tilde_chi_alpha}), whereas $K$ is proportional to (\ref{eq:bar_chi_alpha}). 
The experimental observation that $1/T_1$ is essentially isotropic and that $K$ is anisotropic suggest that
anisotropic effects are not too important for the $p=2k_F$ contributions but do
affect the $p=0$ terms.

Another possible extension of our work concerns of course the two-dimensional systems. Although it is unclear how much of the weak
coupling approach remains valid in higher dimension, our strong
coupling analysis can straightforwardly be applied to higher
dimensional structures. The main difference in that case between the 2D
(or higher) and the 1D study presented here comes from the symmetry of
the various orbitals. In the case of a Cu-O plane, 
in the presence of a Coulomb repulsion
on the oxygen sites ($U_{\mbox{\footnotesize O-2p}} \ne 0$),\cite{unpublished_becker_2}
the related amplitudes for the local processes (\ref{eq:process_AL(a)}) and (\ref{eq:process_AL(c)}) are
not equal anymore, and a cancellation of these terms by symmetry arguments as
assumed by Mila-Rice does not occur. Only the contribution like (\ref{eq:process_AL(b)}) will vanish by
symmetry arguments. The transferred Mila-Rice contributions
(\ref{eq:process_AT(a)}) via the O-2p orbital, which always cost the Coulomb
energy $U_{\mbox{\footnotesize O-2p}}$, and the local processes (\ref{eq:process_AL(a)}) have
exactly the same combinatorial factor and the same amplitude; thus the term
(\ref{eq:process_AL(a)}) cancels out the term (\ref{eq:process_AT(a)}) for the antiferromagnetic wave vector.
This suggests for $U_{\mbox{\footnotesize Cu-3d}}\to \infty$ that the
antiferromagnetic contribution to the relaxation of the copper nuclei via an
isotropic interaction comes from local terms (see (\ref{eq:process_AL(c)}))
and from new transferred combined terms of third order (see (\ref{eq:process_AT(d)})),
 while the transferred contributions proposed up to now are absent (see
(\ref{eq:process_AT(b)}) and (\ref{eq:process_AT(c)})).

\acknowledgments
This work was initiated by a suggestion 
from the late H.J. Schulz whom we wish to acknowledge here.

\begin{appendix}

\section{Anisotropic hyperfine couplings} 
\label{anistropic_hyperfin}

Taking into account anisotropic hyperfine couplings related to the orbitals
Cu-3d and O-2p we have to replace
(\ref{eq:hamN}) by
\begin{equation} \label{eq:hamN_aniso}
H_N^\prime=\sum_{j\varsigma}  
		C_A 		I_{j}^\varsigma	S_{Aj}^\varsigma
	    +	C_a^\varsigma 	I_{j}^\varsigma S_{aj}^\varsigma
            + 	C_B 		J_{j}^\varsigma	S_{Bj}^\varsigma
            + 	C_b^\varsigma 	J_{j}^\varsigma S_{bj}^\varsigma
\end{equation}
The sum on $\varsigma$ is over components of the diagonal hyperfine
tensors $ C_\eta^\varsigma$. 

In the weak interaction limit of the four-band model we can perform the same
calculations as done before, and we will end up with the bosonized expression (\ref{eq:boscomp}), where now
the electron-nuclear interaction is given by 
\begin{eqnarray}
H_N &=&   \sum_\varsigma   \int dx 
	\big( C_A |\lambda_{A\alpha}|^2+ C_a^\varsigma
|\lambda_{a\alpha}|^2 \big)
	I^\varsigma 
	\big(
\bar{S}^\varsigma_{\alpha}+\tilde{S}^\varsigma_{a\alpha}\big)
\nonumber\\
&+&  	  \sum_\varsigma   \int dx 
	\big( C_B |\lambda_{B\alpha}|^2+ C_b^\varsigma
|\lambda_{b\alpha}|^2 \big)
	J^\varsigma
	\big(
\bar{S}^\varsigma_{\alpha}+\tilde{S}^\varsigma_{b\alpha}\big)\;.
\nonumber\\ 
\end{eqnarray}
Here we used the fact that
$\tilde{S}^\varsigma_{A\alpha}=\tilde{S}^\varsigma_{a\alpha}$ and
$\tilde{S}^\varsigma_{B\alpha}=\tilde{S}^\varsigma_{b\alpha}$. In general
the explicit bosonized expressions for the spin part of
$\tilde{S}^x_{\eta\alpha}$ and $\tilde{S}^y_{\eta\alpha}$ in (\ref{eq:szoscillating}) differ from
$\tilde{S}^z_{\eta\alpha}$, but finally for a spin-symmetric model there
will be no influence on the correlation functions. Thus only the
coefficients are slightly modified and vary for the
different directions $\varsigma=x,y,z$. Formally the contributions to the $\varsigma$-directions of
the Knight shifts $K_{Cu}^\varsigma$ and $K_{O}^\varsigma$, as well as
the contributions to the relaxation
times $T_{1,Cu}^\varsigma$ and  $T_{1,O}^\varsigma$ are given by (\ref{eq:K_eta_gamma})
and (\ref{eq:T_1_eta_gamma}) performing the replacements
\begin{eqnarray}
 |\lambda_{A\alpha}|^2 & \to &
  |\lambda_{A\alpha}|^2+\frac{C_a^\varsigma}{C_A} |\lambda_{a\alpha}|^2 
\\
 |\lambda_{B\alpha}|^2 & \to &
  |\lambda_{B\alpha}|^2+\frac{C_b^\varsigma}{C_B} |\lambda_{b\alpha}|^2
\end{eqnarray}
in the expressions of the form factors defined in
Table~\ref{table:formfactors_general_small_U}.

In the strong interaction limit of the four-band model the inclusion of
anisotropic hyperfine interactions results in 
\begin{eqnarray}
H_N^{\prime\prime} &=&   \sum_{j\varsigma}
	 \big( 2 C_A F_{A,L} + C_a^\varsigma F_{a,L} \big) I^\varsigma_j
	 S^\varsigma_{aj}
\nonumber\\
&+&	 \sum_{j\varsigma}  
	F_{A,T} I^\varsigma_j ( S^\varsigma_{a,j-1}+ S^\varsigma_{a,j+1}) 
\nonumber\\
&+&  	  \sum_{j\varsigma}  
	\big( C_B F_{B,T} + C_b^\varsigma F_{b,T}\big)
	J^\varsigma_j
	\big( S^\varsigma_{aj}+ S^\varsigma_{a,j+1} \big)\;.
\end{eqnarray}
Here the new defined parameters which describe the additional couplings to
the local Cu-3d spins are given by  $F_{a,L}=1$ and
$F_{b,T}=t_{ab}^2/(\epsilon_a -\epsilon_b)^2$, whereas all the others were
defined in Section~\ref{strong_interaction_limit}.
For the copper atom the local contribution is modified, whereas for
the oxygen atom it is the transferred one. 

\section{The sine-Gordon model} \label{sine-gordon-model}

At half-filling the spin part as well as the charge part of the
Hamiltonian (\ref{eq:boscomp}) are described by a sine-Gordon
model $H^\nu_{SG}=H^\nu_0+H^\nu_U$ where
\begin{equation}
H^\nu_U =    \frac{2 a U_\nu}{(2\pi a)^2}     \int_0^L dx
                \cos[ \sqrt{8}\Phi_\nu ]
\end{equation}
For this model two different regimes exist depending on the value
of the parameter $K_\nu$. A massive regime ($\nu_m$) for
\begin{equation} \label{eq:massive_inequality}
2\pi u_\nu (K_\nu-1)<|U_\nu|\;,
\end{equation}
where the perturbation of $H_\nu^U$ is relevant, and a massless ($\nu_o$) for
\begin{equation} \label{eq:massless_inequality}
2\pi u_\nu (K_\nu-1)>|U_\nu|\;,
\end{equation}
where the perturbation is irrelevant.

\subsection{Massive regime ($\nu_m$)}

When the cosine term is relevant, the conformal symmetry is lost
and the elementary excitations become massive particles. To
compute the correlation functions we can approximate  the cosine
term by
\begin{equation}
H^\nu_m=\frac{m_\nu ^2}{2} \int_0^L dx
        (\Phi_\nu - \langle \Phi_\nu \rangle )^2\;,
\end{equation}
where the mass can be obtained from the exact solution of the
sine-Gordon equation. For small $U_\nu$ one has
\begin{equation}
m_\nu = \left(   \frac{4 K_\nu |U_\nu | a}{\pi u_\nu}
    \right) ^\frac{1}{2-2 K_\nu} a^{-1}\;.
\end{equation}
This Hamiltonian describes the fluctuations $\delta\Phi_\nu$ of
$\Phi_\nu$ about its mean value $\langle \Phi_\nu \rangle=0 $.
For such a system the Green's function $\langle T_\tau \Phi_\nu
({\bf r}_\nu) \Phi_\nu ({\bf 0})\rangle_{\nu_m} $ of the Laplace operator
defined on the domain $A_\nu=[0 < u_\nu \tau < u_\nu \beta ,0 < x <
L]$ is given by
\begin{equation}\label{eq:greenfkt_m}
G^{\nu_m} (r_\nu )=\frac{K_\nu}{2} {\rm K}_0 \left[ m_\nu (r_\nu +
a)\right]\;.
\end{equation}
$K_0$ is the Bessel function of zero order.
\subsection{Massless regime ($\nu_o$)}
In this regime, the bare parameters are renormalized up to the fixed
point values $u_\nu \to u_\nu^*$, $K_\nu \to K_\nu^*$, and $U_\nu
\to 0$ without changing the basic Luttinger Liquid behavior of the
unperturbed part $H^\nu_0$. For this model, the Green's functions for
the unperturbed part regularized for large distances by $R_\nu$ and
for short distances by the lattice constant $a$ can be expressed as
\begin{equation}\label{eq:greenfkt_o}
G^{\nu_o} (r_\nu)=\frac{K_\nu}{2} \ln \left[ R_\nu /(r_\nu + a) \right]
\end{equation}
or as the following limit
\begin{equation}
G^{\nu_o}(r_\nu) =  \lim_{m_\nu \to 0} G^{\nu_m} (r_\nu)
\end{equation}
\subsection{Correlation functions}
Typical spin-spin correlation functions of the original fermions
defined in (\ref{eq:fermion_as_boson}) are combinations of
exponentials of $\Phi_\nu$. For a Gaussian model these functions
can be expressed in terms of the Green's functions
(\ref{eq:greenfkt_m}) or (\ref{eq:greenfkt_o}) depending on the
phase $\nu_i$,
\begin{eqnarray}
& & \left\langle
    \exp[i\gamma_1 \Phi_\nu(1)]\ldots \exp[i\gamma_N \Phi_\nu(N)]
\right\rangle_{\nu_i}=
\nonumber\\
& & e^{-\sum_{n>m}^N \gamma_n \gamma_m G^{\nu_i}(r_\nu^{nm})}
    e^{-\frac{1}{2}\sum_{n}^N \gamma_n^2 G^{\nu_i}(r_\nu^{nn})}\;.
\end{eqnarray}
\section{Local states in the strong interaction limit}
\label{local_states_for_u_a}
The projected Hamiltonian (\ref{eq:hamcomp}) is expressed as
\begin{equation}
\hat{P}H\hat{P}=\hat{P}(L+K)\hat{P}\;,
\end{equation}
where $L$ denotes the local system, whereas $K$ includes all possible hopping
terms of $H$. The eigenstates of $L$ are given by
\begin{equation}
|n_1,n_2,\ldots,n_j,\ldots,n_N \rangle = \prod_{j=1}^N |n_j\rangle\;,
\end{equation}
where $n_j$ labels the local states $n$ on site $j$. The local states and
energies are shown in Table~\ref{table:local_states}. For simplicity we use the
short notation
\begin{equation}
|0_1,0_2,\ldots,n_j,\ldots,m_i,\ldots0_{N-1},0_N \rangle\equiv |n_j,m_i\rangle
\end{equation}
(local ground state
configurations are labeled by $|0_j\rangle$). The energy of such a state
 is given by
\begin{equation}
E_{n,m}=(N-2)\epsilon_0+\epsilon_n+\epsilon_m\;.
\end{equation}

Now, we can expand the projection operator $\hat{P}$ onto the unperturbed
eigenstates of $L$. Here for the half-filled case, we are only 
interested in the projection $\hat{P}$ onto the
state $|0\rangle=\prod_{j=1}^N |0_j\rangle$ with the energy $E_0=N\epsilon_0$, thus we get
\begin{equation}
\hat{P}=\sum_i \hat{P}^{(i)}\;,
\end{equation}
where the first orders are given by
\begin{eqnarray}
\hat{P}^{(0)}   &=&\hat{P}_P \nonumber
\\
\hat{P}^{(1)}   &=&\hat{P}_{PQ}+\hat{P}_{QP} \nonumber
\\
\hat{P}^{(2)}   &=&\hat{P}_{PQQ}+\hat{P}_{QPQ}+\hat{P}_{QQP}\nonumber
\\
    &-&(\hat{P}_{PPQ^2}+\hat{P}_{PQ^2P}+\hat{P}_{Q^2PP})
\label{eq:P(i)}
\end{eqnarray}
with
\begin{eqnarray}
&\hat{P}_P&=\hat{P}_0 \nonumber
\\
&\hat{P}_{PQ}&=\hat{P}_0K\hat{Q}_0\frac{1}{E_0-L}\hat{Q}_0 \nonumber
\\
&\hat{P}_{QP}&=\hat{Q}_0\frac{1}{E_0-L}\hat{Q}_0K\hat{P}_0 \nonumber
\\
&\hat{P}_{PQQ}&=\hat{P}_0K\hat{Q}_0\frac{1}{E_0-L}\hat{Q}_0K\hat{Q}_0\frac{1}{E_0-L}\hat{Q}_0 \nonumber
\\
&\hat{P}_{QPQ}&=\hat{Q}_0\frac{1}{E_0-L}\hat{Q}_0K\hat{P}_0K\hat{Q}_0\frac{1}{E_0-L}\hat{Q}_0 \nonumber
\\
&\hat{P}_{QQP}&=\hat{Q}_0\frac{1}{E_0-L}\hat{Q}_0K\hat{Q}_0\frac{1}{E_0-L}\hat{Q}_0K\hat{P}_0 \nonumber
\\
&\hat{P}_{PPQ^2}&=\hat{P}_0K\hat{P}_0K\hat{Q}_0\frac{1}{(E_0-L)^2}\hat{Q}_0 \nonumber
\\
&\hat{P}_{PQ^2P}&=\hat{P}_0K\hat{Q}_0\frac{1}{(E_0-L)^2}\hat{Q}_0K\hat{P}_0 \nonumber
\\
&\hat{P}_{Q^2PP}&=\hat{Q}_0\frac{1}{(E_0-L)^2}\hat{Q}_0K\hat{P}_0K\hat{P}_0
\label{eq:P_ppq}
\end{eqnarray}
The projection operator $\hat{Q}_0$ denotes $1-\hat{P}_0$.

\begin{figure}
\centerline{\epsfig{file=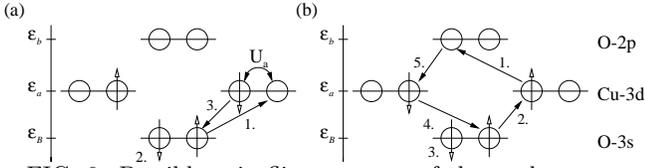,angle=0,width=8.6cm}}
\caption{Possible spin flip processes of the nuclear oxygen spin via a Fermi
contact interaction at half-filling. (a) is
present in the 1D version of the Shastry model,
 whereas (b) is another possible process which
includes some superexchange contributions. At half-filling there are only
transferred contributions (T), local processes (L) are absent. The numbers (1.,2.,3.,4. and
5.) denote the sequence of the intermediate steps. }
\label{fig:process_oxygen}
\end{figure}

\begin{figure}
\centerline{\epsfig{file=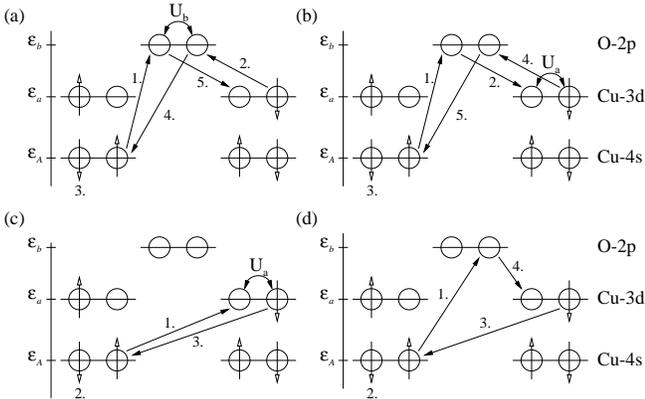,angle=0,width=8.6cm}}
\caption{Possible transferred spin flip processes of the nuclear copper spin
like $\hat{P}S^+_{Aj}\hat{P}$ via a Fermi
contact interaction at half-filling. (a) and (b) appear in the 1D
Mila-Rice model, (c) pertains to the 1D the Shastry model ,
whereas (d) is a combination of both.}
\label{fig:process_copper_trans}
\end{figure}

We can compute the projected electron-nuclear interaction term
$\hat{P}H_N\hat{P}$. The projection affects only the electronic spins, and
we have to evaluate projected local s orbital spin operators such as
$\hat{P}{\bf S}_{\eta j}\hat{P}$.
For example, the second order processes
(see Fig.~\ref{fig:process_oxygen}(a) and \ref{fig:process_copper_trans}(c)) are given by
\begin{equation}\label{eq:S_Aj(2)}
 {\bf S}_{\eta j}^{(2)} =  (\hat{P}^{(0)}+\hat{P}^{(1)})  {\bf S}_{Aj} (\hat{P}^{(0)} + \hat{P}^{(1)})
\end{equation}
Introducing (\ref{eq:P(i)}) and (\ref{eq:P_ppq}) in (\ref{eq:S_Aj(2)}) only
\begin{equation}
 {\bf S}_{Aj}^{(2)} =\hat{P}_{PQ}  {\bf S}_{Aj} \hat{P}_{QP}
\end{equation}
will contribute, because in a half-filled system the first hopping process
brings the system out of the ground state $|0\rangle$ and the second one brings
it back to a possible ground state configuration $|0\rangle$. Here only the projections
\begin{eqnarray}
\hat{P}_{PQ}&=& \frac{t_{Aa}}{E_0-E_{\bar{4},13}}
|0\rangle \langle\bar{4}_j,13_{j\pm 1}|
\\
\hat{P}_{QP}&=& \frac{t_{Aa}}{E_0-E_{\bar{4},13}} |\bar{4}_j,13_{j\pm 1}\rangle \langle 0 |
\end{eqnarray}
could generate finite matrix elements for second order
contributions to the Cu-4s spin ${\bf S}_{Aj}$. For the O-3s spin ${\bf
S}_{Bj}$ it will be
\begin{eqnarray}
\hat{P}_{PQ}&=& \frac{-t_{Ba}}{E_0-E_{12}} |0 \rangle \langle 12_j |
\\
\hat{P}_{QP}&=& \frac{-t_{Ba}}{E_0-E_{12}} |12_j \rangle \langle 0 |
\end{eqnarray}
or
\begin{eqnarray}
\hat{P}_{PQ}&=&  \frac{t_{Ba}}{E_0-E_{\bar{2},13}} |0 \rangle \langle \bar{2}_j,13_{j+1} |
\\
\hat{P}_{QP}&=& \frac{t_{Ba}}{E_0-E_{\bar{2},13}} |\bar{2}_j,13_{j+1}
\rangle \langle 0 | \;.
\end{eqnarray}
Finally, the projected spins ${\bf S}_{\eta j}^{(2)}$ are given by
\begin{eqnarray}
{\bf S}_{Aj}^{(2)} & = & |\lambda_{Aa}|^2 \hat{P}_0 \left( {\bf S}_{a,j-1}+ {\bf S}_{a,j+1} \right)\hat{P}_0
\\
{\bf S}_{Bj}^{(2)} & = & |\lambda_{Ba}|^2 \hat{P}_0 \left( {\bf S}_{a,j}+ {\bf S}_{a,j+1} \right)\hat{P}_0
\;.
\end{eqnarray}
with
\begin{eqnarray}
\lambda_{Aa} &=& \frac{t_{Aa}}{\epsilon_A-\epsilon_a-U_a}
\\
\lambda_{Ba} &=&  \pm \frac{t_{Ba}}{\epsilon_B-\epsilon_a-U_a}\;.
\end{eqnarray}
Higher order contributions could be computed using the same procedure
as for the above examples.
\end{appendix}

%\bibliography{nmr_1D_CuO}
%\bibliographystyle{prsty}

\newpage
\onecolumn

\begin{table}
\narrowtext
\caption{The lowest order temperature dependence of the Knight shifts
and of the relaxation rates measured on the copper site ($A$) or the oxygen
site ($B$) at half-filling (0) and for finite doping ($\delta$). Within a
line, the undefined constants $const.$ are the same for copper and oxygen. Temperature dependences
are given up to logarithmic corrections.}
\begin{tabular}{lll} & Copper & Oxygen \\ \tableline\\ $K_{\eta}^0$ &
$\bar{F}_{A\alpha}\times const.$ & $\bar{F}_{B\alpha}\times const.$\\
$K_{\eta}^\delta$ & $\bar{F}_{A\alpha}\times const.$ &
$\bar{F}_{B\alpha}\times const.$ \\ \\ \tableline \\ $1/\bar{T}_{1\eta}^0$ &  $(\bar{F}_{A\alpha})^2\times const. \times T$ &  $(\bar{F}_{B\alpha})^2\times const. \times T$\\
$1/\tilde{T}_{1\eta}^0$ &  $(\tilde{F}_{A\alpha}^0 )^2\times const.$ &
$(\tilde{F}_{B\alpha}^0 )^2\times const.$\\ \\ \tableline \\
$1/\bar{T}_{1\eta}^\delta$ &  $(\bar{F}_{A\alpha})^2\times const. \times T$
&  $(\bar{F}_{B\alpha})^2 \times const. \times T$ \\
$1/\tilde{T}_{1\eta}^\delta$ &  $(\tilde{F}_{A\alpha}^\delta )^2 \times
const. \times T^{K_c^*}$ &  $(\tilde{F}_{B\alpha}^\delta )^2 \times const.
\times T^{K_c^*}$ \\
\end{tabular}
\label{table:knightshifts_relaxationrates_small_U}
\end{table}

\begin{table}
\widetext
\caption{ $\bar{F}_{\eta\alpha}$ and  $\tilde{F}_{\eta\alpha}^\gamma$ are
the non-oscillating  and the oscillating contributions to the form factors
of the four-band model in the weak interaction limit at half-filling
($\gamma=0$) and for small doping ($\gamma=\delta$) for copper ($A$) and
oxygen ($B$). The coefficients $\lambda_{\eta\alpha}$ come from the
projection onto the lowest band ($\alpha$) for small Coulomb
interaction. $l_\delta$ and $l_{m_c}$ denote the characteristic lengths
related to the doping and the mass for a charge-gap system and $K_c$ is
the Luttinger liquid parameter which controls the anomalous exponents of the
correlation functions. Finally, $K_0$ is the Bessel function of zero order.}
\begin{tabular}{ddd} & Copper & Oxygen\\ \tableline \\$\bar{F}_{\eta\alpha}$  &
$|\lambda_{A\alpha}|^2$ & $|\lambda_{B\alpha}|^2$ \\
$(\tilde{F}_{\eta\alpha}^0)^2$  & $2 |\lambda_{A\alpha}|^4
\left(\frac{a}{l_{m_c}}\right)^{K_c}$ & $0$ \\
$(\tilde{F}_{\eta\alpha}^\delta)^2$ & $2 |\lambda_{A\alpha}|^4 \cosh\left[
K_c K_0\left( \frac{l_\delta}{l_{m_c}}\right) \right]
\left(\frac{a}{l_{m_c}}\right)^{K_c}$ & $2 |\lambda_{B\alpha}|^4\sinh\left[
K_c K_0\left( \frac{l_\delta}{l_{m_c}}\right) \right]
\left(\frac{a}{l_{m_c}}\right)^{K_c}$ \\ \\
\end{tabular}
\label{table:formfactors_general_small_U}
\end{table}

\begin{table}
\widetext
\caption{The non-oscillating and oscillating form
factors at half-filling ($\gamma=0$) and for small doping  ($\gamma=\delta$) for copper and oxygen. $\bar{F}_{\eta\alpha}^{Bu}$ and
$\tilde{F}_{\eta\alpha}^{\gamma Bu}$ denote the form factors obtained for
the weak interaction Bulut model with a Mila-Rice and Shastry like
isotropic electron nuclear interaction term. The coefficients
$\lambda_{\eta\alpha}$ are given by (\ref{eq:proj}) and (\ref{eq:coefproj}).
By contrast, $\lambda_{\eta a}^{Bu}$
is the characteristic coefficient of the Bulut model related to the overlap
between the Cu-4s orbitals (A) the Cu-3d
orbitals (a) or respectively between the O-3s orbitals (B) and the Cu-3d
orbitals (a) performed in real space. The other parameters were explained in
Table~\ref{table:formfactors_general_small_U}.}
\begin{tabular}{ddd} & Copper & Oxygen
\\ \tableline\\ $\bar{F}_{\eta\alpha}^{Bu}$ &
$2|\lambda_{Aa}^{Bu}|^2|\lambda_{a\alpha}|^2$ &
$2|\lambda_{Ba}^{Bu}|^2|\lambda_{a\alpha}|^2$ \\
$(\tilde{F}_{\eta\alpha}^{0Bu} )^2 $   & $ 8
|\lambda_{Aa}^{Bu}|^4|\lambda_{a\alpha}|^4
\left(\frac{a}{l_{m_c}}\right)^{K_c}$ & $ 0 $\\
$(\tilde{F}_{\eta\alpha}^{\delta Bu} )^2$ &
$8|\lambda_{Aa}^{Bu}|^4|\lambda_{a\alpha}|^4\cos(\pi a/ l_\delta )^2
\cosh\left[ K_c K_0\left( \frac{l_\delta}{l_{m_c}}\right) \right]
\left(\frac{a}{l_{m_c}}\right)^{K_c}$ & $8
|\lambda_{Ba}^{Bu}|^4|\lambda_{a\alpha}|^4\sin(\pi a/2l_\delta )^2
\cosh\left[ K_c K_0\left( \frac{l_\delta}{l_{m_c}}\right) \right]
\left(\frac{a}{l_{m_c}}\right)^{K_c}$\\ \\
\end{tabular}
\label{table:formfactors_bulut}
\end{table}

\begin{table}
\narrowtext
\caption{The ratios of the oscil\-lating and non-oscil\-lating contri\-butions to
the oxygen ($\eta=B$) and the copper ($\eta=A$) Knight shifts and relaxation rates at
half-filling ($0$) and away from half-filling ($\delta$)
calculated for the Bulut and the four-band model in the low interaction
limit. The ratios are measured in units of their characteristic overlaps
with the ground state ($|\lambda_{\eta a}^{Bu}||\lambda_{a\alpha}|$ or $|\lambda_{\eta \alpha}|$).
Here the oscillating contributions do
depend on the characteristic length $l_{m_c}$ and $l_\delta$, whereas the
non-oscillating contributions do not.}
\begin{tabular}{ccc} 
& Bulut model& Four-band model \\ \tableline\\
$\frac{K^{0,\delta}_B}{K^{0,\delta}_A}$ & $1$ & $1$ \\ \\ \tableline \\ 
$\frac{1/\bar{T}^{0,\delta}_{1B}}{1/\bar{T}^{0,\delta}_{1A}}$
& $1$ & $1$ \\ \\ \tableline \\ $\frac{1/\tilde{T}^0_{1B}}{1/\tilde{T}^0_{1A}}$ & $0$ & $0$ \\ \\ $\frac{1/\tilde{T}^\delta_{1B}}{1/\tilde{T}^\delta_{1A}}$
& $\frac{\sin(\pi a/2l_\delta)^2}{\cos(\pi a/l_\delta)^2}$
& $ \tanh\left[ K_c K_0\left( \frac{l_\delta}{l_{m_c}} \right)\right]$\\ \\
\end{tabular}
\label{table:ratios}
\end{table}

\begin{table}
\narrowtext
\caption{Possible local states per unit cell of the four band Hubbard
model. The degeneracy is related to the possible spin configurations for a
local state labeled by $|n_j \rangle$.}
\begin{tabular}{dlll} 
degeneracy & local energies & state \\ \tableline 2
&   $\epsilon_{\bar{7}}=\epsilon_A +2\epsilon_B$
&   $|\bar{7}_j\rangle$\\ 2
&   $\epsilon_{\bar{6}}=2\epsilon_A +\epsilon_B$
&   $|\bar{6}_j\rangle$\\ 1
&   $\epsilon_{\bar{5}}=2\epsilon_A +2\epsilon_B$
&   $|\bar{5}_j\rangle$\\ 4
&   $\epsilon_{\bar{4}}=\epsilon_A +2\epsilon_B+\epsilon_a$
&   $|\bar{4}_j\rangle$\\ 4
&   $\epsilon_{\bar{3}}=\epsilon_A +2\epsilon_B+\epsilon_b$
&   $|\bar{3}_j\rangle$\\ 4
&   $\epsilon_{\bar{2}}=2\epsilon_A +\epsilon_B+\epsilon_a$
&   $|\bar{2}_j\rangle$\\ 4
&   $\epsilon_{\bar{1}}=2\epsilon_A +\epsilon_B+\epsilon_b$
&   $|\bar{1}_j\rangle$\\ 2
&   $\epsilon_{0}=2\epsilon_A +2\epsilon_B+\epsilon_a$
&   $|0_j\rangle$ \\ 2
&   $\epsilon_{1}=2\epsilon_A +2\epsilon_B+\epsilon_b$
&   $|1_j\rangle$ \\ 8
&   $\epsilon_{2}=\epsilon_A +2\epsilon_B+\epsilon_a+\epsilon_b$
&   $|2_j\rangle$ \\ 2
&   $\epsilon_{3}=\epsilon_A +2\epsilon_B+2\epsilon_b+U_b$
&   $|3_j\rangle$\\ 8
&   $\epsilon_{4}=2\epsilon_A +\epsilon_B+\epsilon_a+\epsilon_b$
&   $|4_j\rangle$\\ 2
&   $\epsilon_{5}=2\epsilon_A +\epsilon_B+2\epsilon_b+U_b$
&   $|5_j\rangle$\\ 4
&   $\epsilon_{6}=2\epsilon_A +2\epsilon_B+\epsilon_a+\epsilon_b$
&   $|6_j\rangle$\\ 4
&   $\epsilon_{7}=\epsilon_A +2\epsilon_B+\epsilon_a+2\epsilon_b+U_b$
&   $|7_j\rangle$\\ 1
&   $\epsilon_{8}=2\epsilon_A +2\epsilon_B+2\epsilon_b+U_b$
&   $|8_j\rangle$\\ 4
&   $\epsilon_{9}=2\epsilon_A +\epsilon_B+\epsilon_a+2\epsilon_b+U_b$
&   $|9_j\rangle$\\ 2
&   $\epsilon_{10}=2\epsilon_A +2\epsilon_B+\epsilon_a+2\epsilon_b+U_b$
&   $|10_j\rangle$\\ 2
&   $\epsilon_{11}=\epsilon_A +2\epsilon_B+2\epsilon_a+U_a$
&   $|11_j\rangle$\\ 2
&   $\epsilon_{12}=2\epsilon_A +\epsilon_B+2\epsilon_a+U_a$
&   $|12_j\rangle$\\ 1
&   $\epsilon_{13}=2\epsilon_A +2\epsilon_B+2\epsilon_a+U_a$
&   $|13_j\rangle$\\ 4
&   $\epsilon_{14}=\epsilon_A +2\epsilon_B+2\epsilon_a+U_a+\epsilon_b$
&   $|14_j\rangle$\\ 4
&   $\epsilon_{15}=2\epsilon_A +\epsilon_B+2\epsilon_a+U_a+\epsilon_b$
&   $|15_j\rangle$\\ 2
&   $\epsilon_{16}=2\epsilon_A +2\epsilon_B+2\epsilon_a+U_a+\epsilon_b$
&   $|16_j\rangle$\\ 2
&   $\epsilon_{17}=\epsilon_A +2\epsilon_B+2\epsilon_a+U_a+2\epsilon_b+U_b$
&   $|17_j\rangle$\\ 2
&   $\epsilon_{18}=2\epsilon_A +\epsilon_B+2\epsilon_a+U_a+2\epsilon_b+U_b$
&   $|18_j\rangle$\\
1&   $\epsilon_{19}=2\epsilon_A +2\epsilon_B+2\epsilon_a+U_a+2\epsilon_b+U_b$
&   $|19_j\rangle$\\
\end{tabular}
\label{table:local_states}
\end{table}

\newpage 

\end{document}